\documentclass[aip,jcp,reprint,a4paper]{revtex4-1}
\pdfoutput=1

\usepackage[utf8]{inputenc}
\usepackage{graphicx}
\usepackage{amsmath,amssymb}
\usepackage{calc}
\usepackage{multirow}
\usepackage[colorlinks=false]{hyperref}
\usepackage{subfigure}
\usepackage{tikz}
\usetikzlibrary{decorations,arrows,shapes}

\newcommand{\refeq}[1]{Eq.\,(\ref{#1})}
\newcommand{\refeqq}[2]{Eqs.\,(\ref{#1},\ref{#2})}
\newcommand{\refsec}[1]{Sec.\,\ref{#1}}
\newcommand{\refapp}[1]{Appendix \ref{#1}}
\newcommand{\reffig}[1]{Fig.\,\ref{#1}}
\newcommand{\refcite}[1]{Ref.\,\onlinecite{#1}}

\newcommand{\topn}{\varnothing}
\newcommand{\prim}[1]{{[#1]}}
\newcommand{\wo}{\!\setminus\!}
\newcommand{\multidxskip}[3]{{#1}^{{#2}\wo{#3}}}
\newcommand{\multidxrepl}[4]{{#1}^{{#2}\wo{#3}}_{\phantom{{#2}\wo}{#4}}}
\newcommand{\bbar}[1]{\bar{\bar{#1}}}

\begin{document}

\title{Multi-Layer Potfit: An Accurate Potential Representation for Efficient High-Dimensional Quantum Dynamics}

\author{Frank Otto}
\email[]{E-mail: Frank.Otto@pci.uni-heidelberg.de}
\affiliation{Theoretische Chemie, Universit\"at Heidelberg,
    Im Neuenheimer Feld 229, 69120 Heidelberg, Germany}

\date{\today}


\begin{abstract}

The multi-layer multi-configuration time-dependent Hartree method
(ML-MCTDH) is a highly efficient scheme for studying the dynamics of
high-dimensional quantum systems. Its use is greatly facilitated if
the Hamiltonian of the system possesses a particular structure 
through which the multi-dimensional matrix elements can be computed
efficiently. In the field of quantum molecular dynamics, the effective
interaction between the atoms is often described by potential energy surfaces
(PES), and it is necessary to fit such PES into the desired structure. For
high-dimensional systems, the current approaches for this fitting process
either lead to fits that are too large to be practical, or their accuracy is
difficult to predict and control.

This article introduces \emph{multi-layer Potfit} (MLPF),
a novel fitting scheme that results in a PES representation in
the hierarchical tensor (HT) format. The scheme is based on the
hierarchical singular value decomposition, which can yield a near-optimal
fit and give strict bounds for the obtained accuracy.
Here, a recursive scheme for using the HT-format PES within ML-MCTDH
is derived, and theoretical estimates as well as a computational example
show that the use of MLPF can reduce the numerical effort for ML-MCTDH
by orders of magnitude, compared to the traditionally used \textsc{Potfit}
representation of the PES. Moreover, it is shown that MLPF is especially
beneficial for high-accuracy PES representations, and it turns out that
MLPF leads to computational savings already for comparatively small
systems with just four modes.

\end{abstract}

\maketitle

\section{Introduction}

Still today, the accurate numerical treatment of quantum systems with many
degrees of freedom (DOFs) is a challenging task. Just to represent the wavefunction,
most approaches require an amount of data which scales exponentially with
the number of dimensions, a situation which has become known as the
\emph{curse of dimensionality}\cite{bel61}. Worse, the quantum-mechanical
expressions involve high-dimensional integrations (e.g. for computing matrix
elements of operators), and sophisticated methods are necessary for
treating this \emph{quadrature problem}.

In recent years, progress on these problems has been made on multiple fronts.
Sparse grid integration methods, as pioneered by Smolyak\cite{smo63:240},
have been used by Ávila and Carrington as well as
by Lauvergnat and Nauts to compute vibrational spectra
of 12D systems\cite{avi11:054126,avi11:064101,lau13:xxx}. Another approach, which
is also taken in the present work, \emph{formally} employs a full product grid
but uses tensor decomposition techniques to represent wavefunctions and
operators in a compact manner. Perhaps the most successful exponent of this approach is the
\emph{multi-configuration time-dependent Hartree} method (MCTDH)\cite{mey90:73,bec00:1,mey09:book,mey12:351}
which employs a \emph{Tucker decomposition}\cite{tuc63:122} for the wavefunction.
If the Hamiltonian is suitably structured (more on that below), MCTDH can
be used to treat realistic systems up to around 20D%
\cite{raa99:936,cat01:2088,mar05:204310,ven07:184303,ven09:034308,sch11:234307},
and even more if the system is weakly correlated or if low accuracy is sufficient%
\cite{nes03:24,wan00:9948}. The \emph{multi-layer} generalization of the MCTDH
scheme (ML-MCTDH), first formulated by Wang and Thoss\cite{wan03:1289} and
later reformulated by Manthe\cite{man08:164116} as a recursive algorithm
for arbitrary layering schemes, has already been used to treat systems with hundreds of DOFs%
\cite{wan03:1289,cra07:144503,wan10:78,wes11:184102,cra11:064504,ven11:044135,bor12:751}.
ML-MCTDH uses a tensor decomposition which has then become known in the mathematical
literature as \emph{hierarchical tensor} or \emph{hierarchical Tucker} (HT) format\cite{hac09:706},
and developing techniques for operating on tensors in HT- and related formats is
an active field of research. The monograph by Hackbusch\cite{hac12} gives an
in-depth overview of the state of the art, and may be supplemented by the
recent literature survey in \refcite{gra13:survey}.

To overcome the quadrature problem, MCTDH requires that the Hamiltonian
can be expressed as a sum of products of one- (or low-)dimensional operators.
For the kinetic energy part of the Hamiltonian, this can readily be achieved by
using a suitable coordinate system. Most notably, the polyspherical approach
of Gatti et al.\cite{gat09:1} can yield an exact kinetic energy operator for arbitrary
system sizes, and a software package (TANA) is available for carrying out this
procedure automatically\cite{ndo12:034107}.
For the potential energy part of the Hamiltonian, in general only model systems
possess the desired sum-of-products form.  But in quantum molecular dynamics,
the accurate treatment of a realistic system requires the use of a \emph{potential energy surface} (PES),
which is usually obtained as a complicated analytical fit to a large set of electronic
structure calculations. For spectroscopic applications, it may be sufficient to use
normal modes and do a Taylor expansion of the PES, as in the vibronic
coupling Hamiltonian model\cite{koe84:59} for which ML-MCTDH studies with
up to 66D\cite{men12:134302,men13:014313} have been carried out. But a
Taylor expansion is only \emph{locally} a good fit to the PES, whereas for many applications
(e.g. floppy molecules, or scattering)  one needs a more \emph{globally} good fit.
The next paragraph will review several approaches for finding such a fit.
But first it should be noted that the quadrature problem may
alternatively be solved by employing time-dependent grids, as in
Manthe's CDVR method\cite{man96:6989} which also works in conjunction with
ML-MCTDH\cite{man09:054109}, where it has been successfully used to treat
a 189D system-bath problem which involved a 9D PES\cite{wes11:184102}
as well as systems with global PES up to 21D%
\cite{ham11:224305,ham12:054105,wod12:11249,wes13:014309,wel12:244106}.
The drawback of this approach is that it requires an extreme amount of PES
function evaluations, which can easily become a bottleneck for the calculation,
and massively parallel architectures (like GPUs) may be needed to overcome
this problem\cite{wel13:164118}.

For high-dimensional systems, perhaps the currently most widely used method for fitting a PES into
sum-of-products form is the $n$-mode representation ($n$-MR, also known as cut-HDMR)
in which the PES is expressed as a sum of terms which each depend only on a
subset of the coordinates. Expressing each such low-dimensional term as a sum
of products is then a much easier task. MCTDH studies using this approach have
been carried out for the Zundel cation\cite{ven07:184302,ven09:234305} and for
malonaldehyde\cite{sch11:234307}. However, great care and effort is required to
obtain a suitable $n$-MR because the method is not variational, in the sense that
adding more terms to the expansion is not guaranteed to lead to a more accurate
potential representation. Another approach for obtaining a sum-of-products fit
is the use of neural networks with exponential activation functions\cite{man06:194105,pra13:6925},
though it has not been demonstrated that this approach can scale to higher dimensions.
From the perspective of tensor decompositions,
the sum-of-products form corresponds to the so-called
\emph{canonical} or \emph{parallel factors} or shortly \emph{CP} tensor format. In this
context, finding a sum-of-products fit means that,
given a general higher-order tensor, one would wish to find an accurate CP
approximation to it, but the number of summands (the \emph{CP-rank}) should not
be too high. It turns out that this is a difficult problem, due to the fact that the
set of CP-tensors with fixed CP-rank is not closed\cite{sil08:1084}. Nevertheless,
some iterative procedures for solving this approximation problem exist (see
Chapter 9.5 in \refcite{hac12} for an overview), but their accuracy is hard to control
as they may get stuck in local minima. The situation improves if one imposes
additional structural constraints on the CP format, specifically, turning to the
Tucker format (i.e. the format used by the MCTDH wavefunction) leads to the
\textsc{Potfit} algorithm\cite{jae96:7974} which has later been re-derived
and named \emph{higher-order singular value decomposition}\cite{lat00:1253} (HOSVD).
This method allows the efficient construction of near-optimal fits with controlled
accuracy, and it is variational, i.e. the accuracy will increase if more terms are added.
Due to this precise and predictable control of accuracy, Potfit is the method of choice for
systems where it is affordable, and hence it has been 
used in numerous MCTDH studies to date. 

Potfit suffers from two limitations. The first limitation is that the algorithm starts
from a full-grid representation of the potential, and hence it is subject to the curse
of dimensionality. Recently, Peláez and Meyer introduced the \emph{multigrid Potfit}
method (MGPF)\cite{pel13:014108} which addresses this limitation by employing two
nested grids, a coarse one and a fine one. This can reduce the amount of data 
that needs to be processed by orders of magnitude, so that it may be feasible
to treat systems with up to about 12D with MGPF.
The second limitation of Potfit is that the number of terms in the Tucker
representation still grows exponentially with dimensionality, which negatively
affects the performance of (ML-)MCTDH, since their numerical
effort depends linearly on this number of terms. Essentially, this problem has
so far precluded the wider use of ML-MCTDH for large systems with general PES
(except for the studies using CDVR).

The present article addresses this second limitation
by employing the \emph{hierarchical} Tucker format for representing
the potential. In contrast to Potfit, this does not result in a sum-of-products
structure for the potential energy operator, but in a hierarchical multi-layer structure,
which motivates the name \emph{multi-layer Potfit} (MLPF) for this new fitting procedure.
It will be shown that this multi-layer operator structure,
when used in conjunction with ML-MCTDH, leads to a solution
of the quadrature problem which is vastly superior to the regular Potfit
approach, in that it can reduce the numerical effort for ML-MCTDH by orders of
magnitude. Additionally, MLPF preserves the variational nature of the
fitting process, and very high-accuracy representations of the PES become feasible.

This article is organized as follows.
Section \ref{sec:mlmctdh} reviews the ML-MCTDH scheme.
In Section \ref{sec:mlop+mlmctdh}, multi-layer operators are
introduced, and the scheme for using them within ML-MCTDH
is derived.
Section \ref{sec:pf+mlpf} presents MLPF as a multi-layer
generalization of Potfit.
Section \ref{sec:discuss} discusses how the computational cost for
ML-MCTDH is reduced by using MLPF, and points out some
limitations of the new scheme.
In Section \ref{sec:compex}, a computational example shows the
actual benefits of MLPF.
Section \ref{sec:summary} concludes.
Two appendices describe the more technical details
of MLPF and the computational cost analysis.

\section{Review of ML-MCTDH}
\label{sec:mlmctdh}

The treatment of ML-MCTDH used in this article closely follows the
approach of Manthe\cite{man08:164116} and its discussion in
\refcite{ven11:044135}. However,
the introduction of a new representation for the potential energy operator
will require changes to the practical equations of motion for ML-MCTDH.
Due to the increased complexity of these equations, it is beneficial to
introduce a more compact notation for the quantities involved. In the
recursive spirit of the resulting expressions, this new notation focuses
on the \emph{relative} position of a logical mode
in the multi-layer tree, and keeping explicit track of the layer to which a mode
belongs is not necessary. For completeness and consistency, this section
presents Manthe's ideas and equations in the new notation.

\subsection{Wavefunction ansatz and notation}

A quantum system with $d$ distinguishable degrees of freedom (DOFs)
$q_1, q_2, \ldots, q_d$ is described
by a wavefunction $\psi$ which depends on these coordinates (and on time $t$).
To treat this system computationally, one can employ for each DOF $f$
a set of $N_f$ time-independent orthonormal basis functions $\chi^\prim{f}_\alpha(q_f)$
(henceforth called \emph{primitive basis functions}),
and expand $\psi$ in the resulting product basis:
\begin{align}
&\psi(q_1, \ldots, q_d, t) \nonumber\\
&\quad = \sum_{\alpha_1=1}^{N_1} \cdots \sum_{\alpha_d=1}^{N_d}
B_{\alpha_1 \cdots \alpha_d}(t) \chi^\prim{1}_{\alpha_1}(q_1) \cdots \chi^\prim{d}_{\alpha_d}(q_d)
\label{eq:stdmethod}
\end{align}
The problem with this \emph{standard method} is that the size of $B$ scales exponentially
with $d$, namely $\#B = \prod_{f=1}^d N_f = N^d$ (where $N$ is the geometric mean of the $N_f$).
One approach for solving this problem is the Multi-Configuration Time-Dependent Hartree (MCTDH)
method\cite{mey90:73,bec00:1,mey09:book,mey12:351} which expands $\psi$ in smaller sets
of \emph{time-dependent} orthonormal basis functions $\varphi^\prim{f}_i(q_f,t)$, $i=1 \ldots n_f$. These
\emph{single-particle functions} (SPFs) are in turn expanded in the primitive bases:
\begin{align}
\psi(q_1, \ldots, q_d, t) &=
    \sum_{i_1=1}^{n_1} \cdots \sum_{i_d=1}^{n_d} A_{i_1 \cdots i_d} (t)
    \nonumber\\&\qquad\times
    \varphi^\prim{1}_{i_1}(q_1,t) \cdots \varphi^\prim{d}_{i_d}(q_d,t)
\\
\varphi^\prim{f}_{i}(q_f,t) &= \sum_{\alpha=1}^{N_f} A^\prim{f}_{i;\alpha}(t) \chi^\prim{f}_\alpha(q_f)
    \:\:;\: f = 1 \ldots d
\end{align}
This representation of $\psi$ requires $\sim n^d + dnN$ storage
(where $n$ is a mean value of the $n_f$)  for $A$ and the $A^\prim{f}$,
which is much lower than the storage requirement for $B$ if $n \ll N$. By experience, $n$ can
be smaller than $N$ by a factor between 2 and 10, depending on the system and the required
accuracy of the representation.

Yet, MCTDH does not get rid of the curse of dimensionality, it just alleviates it by lowering
the base of the exponential scaling. Further computational savings can be gained by grouping
DOFs together into logical \emph{modes} $Q_\kappa$, $\kappa=1 \ldots p$, such that the
$\kappa$-th mode contains the DOFs $a_\kappa$ through $b_\kappa$,
and then employing multi-dimensional SPFs:
\begin{align}
Q_\kappa &= (q_{a_\kappa}, \ldots, q_{b_\kappa})
    \:\:;\: \kappa=1 \ldots p
\\
\psi(Q_1, \ldots, Q_p, t) &=
    \sum_{i_1=1}^{\tilde{n}_1} \cdots \sum_{i_p=1}^{\tilde{n}_p} A_{i_1 \cdots i_p} (t)
    \nonumber\\&\qquad\times
    \varphi^{(1)}_{i_1}(Q_1,t) \cdots \varphi^{(p)}_{i_p}(Q_p,t)
\\
\varphi^{(\kappa)}_{i}(Q_\kappa, t) &=
    \sum_{\alpha_1=1}^{N_{a_\kappa}} \cdots \sum_{\alpha_{d_\kappa}=1}^{N_{b_\kappa}}
    A^{(\kappa)}_{i;\alpha_1 \cdots \alpha_{d_\kappa}}(t)
    \nonumber\\&\qquad\times
    \chi^\prim{a_\kappa}_{\alpha_1}(q_{a_\kappa}) \cdots  \chi^\prim{b_\kappa}_{\alpha_{d_\kappa}}(q_{b_\kappa})
    \\
    &\qquad ( i=1 \ldots \tilde{n}_\kappa
    \quad;\quad d_\kappa := b_\kappa - a_\kappa + 1 )
    \nonumber
\end{align}
This \emph{mode combination} reduces the storage requirements to
$\sim \tilde{n}^p + p \tilde{n} N^{d/p}$. As a rule of thumb, the number of
multi-dimensional SPFs that is necessary to keep the same level of accuracy as
when using one-dimensional SPFs, is $\tilde{n}_\kappa \approx n_{a_\kappa} + \cdots + n_{b_\kappa}$,
or in short $\tilde{n} \approx nd/p$.  With this rule, now also the exponent of
the exponential scaling rule is reduced, which further alleviates the curse of
dimensionality.  However, in practice one can group only $d_\kappa=2 \ldots 4$ DOFs
into one mode, because otherwise the effort for treating the multi-dimensional SPFs
($\sim N^{d_\kappa}$) becomes prohibitive. 

But the computational effort can be further reduced by introducing another layer
of basis functions and expansion coefficients. In this scheme, one first groups
the DOFs into a moderate number of modes (say, $p=2 \ldots 4$), and then
splits each mode $Q_\kappa$ into sub-modes $Q_{\kappa,1} , \ldots, Q_{\kappa,p_\kappa}$.
Then one introduces sub-SPFs $\varphi^{(\kappa,\lambda)}_j(Q_{\kappa,\lambda},t)$
($j=1 \ldots n_{\kappa,\lambda}$) and
expands the SPFs $\varphi^{(\kappa)}_i(Q_{\kappa},t)$ in products of these sub-SPFs
(now dropping the tilde from the SPF numbers):
\begin{align}
\psi(Q_1, \ldots, Q_p, t) &=
    \sum_{i_1=1}^{n_1} \cdots \sum_{i_p=1}^{n_p} A_{i_1 \cdots i_p} (t)\,
    \nonumber\\&\qquad\times
    \varphi^{(1)}_{i_1}(Q_1,t) \cdots \varphi^{(p)}_{i_p}(Q_p,t)
\label{eq:totwf}
\end{align}
\begin{align}
\varphi^{(\kappa)}_i (Q_{\kappa},t) &=
    \sum_{j_1=1}^{n_{\kappa,1}} \cdots \sum_{j_{p_\kappa}=1}^{n_{\kappa,p_\kappa}}
    A^{(\kappa)}_{i;j_1 \cdots j_{p_\kappa}}\!(t)\,
    \nonumber\\&\qquad\times
    \varphi^{(\kappa,1)}_{j_1}(Q_{\kappa,1},t) \cdots \varphi^{(\kappa,p_\kappa)}_{j_{p_\kappa}}(Q_{\kappa,p_\kappa},t)
\label{eq:l1spf}
\end{align}
The sub-SPFs themselves are either expanded in the primitive bases,
\begin{align}
\varphi^{(\kappa,\lambda)}_j(Q_{\kappa,\lambda},t) &=
    \sum_{\alpha_1=1}^{N_{a_{\kappa,\lambda}}} \cdots \sum_{\alpha_{d_{\kappa,\lambda}}=1}^{N_{b_{\kappa,\lambda}}}
    A^{(\kappa,\lambda)}_{j;\alpha_1 \cdots \alpha_{d_{\kappa,\lambda}}}\!(t)\,
    \nonumber\\&\qquad\times
    \chi^\prim{a_{\kappa,\lambda}}_{\alpha_1}(q_{a_{\kappa,\lambda}}) \cdots
    \chi^\prim{b_{\kappa,\lambda}}_{\alpha_{d_{\kappa,\lambda}}}(q_{b_{\kappa,\lambda}})
\:,
\end{align}
or, if the mode $Q_{\kappa,\lambda}$ is too big to do so, it is expanded in another
set of sub-sub-SPFs which depend on sub-sub-modes $Q_{\kappa,\lambda,\mu}$, 
and so on.  This \emph{multi-layer MCTDH} (ML-MCTDH) representation of $\psi$
is very flexible and rather compact (in terms of the total number of expansion coefficients).
Manthe estimates\cite{man08:164116} that, under moderate assumptions, the storage
requirements for an ML-MCTDH wavefunction scale polynomially in $d$ ($d^3$ seems
a realistic value).

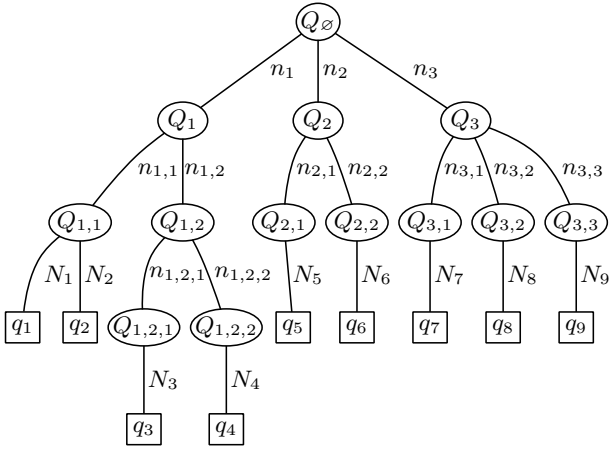
\begin{figure}
{\fontsize{9}{11}\selectfont
\begin{tikzpicture}[>=latex,line join=bevel, scale=0.58, every node/.style={scale=1.0}]
  \pgfsetlinewidth{0.58bp}
\pgfsetcolor{black}
  \draw [] (201bp,221.74bp) .. controls (201bp,233.29bp) and (201bp,250.81bp)  .. (201bp,262.34bp);
  \definecolor{strokecol}{rgb}{0.0,0.0,0.0};
  \pgfsetstrokecolor{strokecol}
  \draw (212.5bp,242bp) node {$n_2$};
  \draw [] (11.756bp,86.318bp) .. controls (12.849bp,95.899bp) and (15.516bp,110.26bp)  .. (22bp,121bp) .. controls (25.302bp,126.47bp) and (30.327bp,131.35bp)  .. (35.082bp,135.2bp);
  \draw (34.5bp,110bp) node {$N_1$};
  \draw [] (317.04bp,155.48bp) .. controls (314.67bp,164.23bp) and (311.06bp,176.53bp)  .. (307bp,187bp) .. controls (305.46bp,190.96bp) and (303.53bp,195.19bp)  .. (301.72bp,198.92bp);
  \draw (327.5bp,177bp) node {$n_{3,2}$};
  \draw [] (362.85bp,155.48bp) .. controls (358.95bp,164.93bp) and (352.31bp,178.18bp)  .. (343bp,187bp) .. controls (333.9bp,195.62bp) and (320.95bp,201.63bp)  .. (310.95bp,205.34bp);
  \draw (371.5bp,177bp) node {$n_{3,3}$};
  \draw [] (222.72bp,155.46bp) .. controls (220.1bp,164.21bp) and (216.2bp,176.5bp)  .. (212bp,187bp) .. controls (210.42bp,190.95bp) and (208.48bp,195.17bp)  .. (206.67bp,198.9bp);
  \draw (233.5bp,177bp) node {$n_{2,2}$};
  \draw [] (179.59bp,155.66bp) .. controls (180.34bp,164.51bp) and (182.07bp,176.84bp)  .. (186bp,187bp) .. controls (187.72bp,191.45bp) and (190.39bp,195.93bp)  .. (193bp,199.74bp);
  \draw (200.5bp,177bp) node {$n_{2,1}$};
  \draw [] (114bp,155.8bp) .. controls (114bp,167.81bp) and (114bp,186.33bp)  .. (114bp,198.3bp);
  \draw (128.5bp,177bp) node {$n_{1,2}$};
  \draw [] (125.37bp,218.37bp) .. controls (142.32bp,230.84bp) and (173.98bp,254.12bp)  .. (190.44bp,266.23bp);
  \draw (178.5bp,242bp) node {$n_1$};
  \draw [] (367bp,86.034bp) .. controls (367bp,98.299bp) and (367bp,119.03bp)  .. (367bp,132.09bp);
  \draw (379.5bp,110bp) node {$N_9$};
  \draw [] (273bp,86.034bp) .. controls (273bp,98.299bp) and (273bp,119.03bp)  .. (273bp,132.09bp);
  \draw (285.5bp,110bp) node {$N_7$};
  \draw [] (184.11bp,86.034bp) .. controls (183.03bp,98.299bp) and (181.2bp,119.03bp)  .. (180.05bp,132.09bp);
  \draw (194.5bp,110bp) node {$N_5$};
  \draw [] (283.97bp,218.11bp) .. controls (265.39bp,230.62bp) and (230.01bp,254.45bp)  .. (212.05bp,266.55bp);
  \draw (270.5bp,242bp) node {$n_3$};
  \draw [] (226bp,86.034bp) .. controls (226bp,98.299bp) and (226bp,119.03bp)  .. (226bp,132.09bp);
  \draw (238.5bp,110bp) node {$N_6$};
  \draw [] (55.981bp,154.82bp) .. controls (62.907bp,163.87bp) and (73.454bp,176.9bp)  .. (84bp,187bp) .. controls (89.687bp,192.45bp) and (96.634bp,197.84bp)  .. (102.41bp,202.04bp);
  \draw (98.5bp,177bp) node {$n_{1,1}$};
  \draw [] (48bp,86.034bp) .. controls (48bp,98.299bp) and (48bp,119.03bp)  .. (48bp,132.09bp);
  \draw (60.5bp,110bp) node {$N_2$};
  \draw [] (88.04bp,87.596bp) .. controls (87.72bp,96.994bp) and (88.327bp,110.37bp)  .. (93bp,121bp) .. controls (95.177bp,125.95bp) and (98.857bp,130.54bp)  .. (102.53bp,134.3bp);
  \draw (111bp,110bp) node {$n_{1,2,1}$};
  \draw [] (89bp,20.024bp) .. controls (89bp,31.852bp) and (89bp,51.498bp)  .. (89bp,64.091bp);
  \draw (101.5bp,42bp) node {$N_3$};
  \draw [] (142bp,20.024bp) .. controls (142bp,31.852bp) and (142bp,51.498bp)  .. (142bp,64.091bp);
  \draw (154.5bp,42bp) node {$N_4$};
  \draw [] (138.36bp,87.639bp) .. controls (135.36bp,96.827bp) and (130.84bp,109.9bp)  .. (126bp,121bp) .. controls (124.27bp,124.96bp) and (122.15bp,129.19bp)  .. (120.18bp,132.92bp);
  \draw (153bp,110bp) node {$n_{1,2,2}$};
  \draw [] (320bp,86.034bp) .. controls (320bp,98.299bp) and (320bp,119.03bp)  .. (320bp,132.09bp);
  \draw (332.5bp,110bp) node {$N_8$};
  \draw [] (273.92bp,155.66bp) .. controls (274.91bp,164.51bp) and (276.93bp,176.85bp)  .. (281bp,187bp) .. controls (282.77bp,191.43bp) and (285.45bp,195.9bp)  .. (288.05bp,199.72bp);
  \draw (295.5bp,177bp) node {$n_{3,1}$};
\begin{scope}
  \definecolor{strokecol}{rgb}{0.0,0.0,0.0};
  \pgfsetstrokecolor{strokecol}
  \draw [rounded corners] (114bp,144bp) ellipse (20bp and 12bp);
  \draw (114bp,144bp) node {$Q_{1,2}$};
\end{scope}
\begin{scope}
  \definecolor{strokecol}{rgb}{0.0,0.0,0.0};
  \pgfsetstrokecolor{strokecol}
  \draw (22bp,86bp) -- (0bp,86bp) -- (0bp,66bp) -- (22bp,66bp) -- cycle;
  \draw (11bp,76bp) node {$q_1$};
\end{scope}
\begin{scope}
  \definecolor{strokecol}{rgb}{0.0,0.0,0.0};
  \pgfsetstrokecolor{strokecol}
  \draw (59bp,86bp) -- (37bp,86bp) -- (37bp,66bp) -- (59bp,66bp) -- cycle;
  \draw (48bp,76bp) node {$q_2$};
\end{scope}
\begin{scope}
  \definecolor{strokecol}{rgb}{0.0,0.0,0.0};
  \pgfsetstrokecolor{strokecol}
  \draw [rounded corners] (48bp,144bp) ellipse (20bp and 12bp);
  \draw (48bp,144bp) node {$Q_{1,1}$};
\end{scope}
\begin{scope}
  \definecolor{strokecol}{rgb}{0.0,0.0,0.0};
  \pgfsetstrokecolor{strokecol}
  \draw (153bp,20bp) -- (131bp,20bp) -- (131bp,0bp) -- (153bp,0bp) -- cycle;
  \draw (142bp,10bp) node {$q_4$};
\end{scope}
\begin{scope}
  \definecolor{strokecol}{rgb}{0.0,0.0,0.0};
  \pgfsetstrokecolor{strokecol}
  \draw (196bp,86bp) -- (174bp,86bp) -- (174bp,66bp) -- (196bp,66bp) -- cycle;
  \draw (185bp,76bp) node {$q_5$};
\end{scope}
\begin{scope}
  \definecolor{strokecol}{rgb}{0.0,0.0,0.0};
  \pgfsetstrokecolor{strokecol}
  \draw (237bp,86bp) -- (215bp,86bp) -- (215bp,66bp) -- (237bp,66bp) -- cycle;
  \draw (226bp,76bp) node {$q_6$};
\end{scope}
\begin{scope}
  \definecolor{strokecol}{rgb}{0.0,0.0,0.0};
  \pgfsetstrokecolor{strokecol}
  \draw (284bp,86bp) -- (262bp,86bp) -- (262bp,66bp) -- (284bp,66bp) -- cycle;
  \draw (273bp,76bp) node {$q_7$};
\end{scope}
\begin{scope}
  \definecolor{strokecol}{rgb}{0.0,0.0,0.0};
  \pgfsetstrokecolor{strokecol}
  \draw (331bp,86bp) -- (309bp,86bp) -- (309bp,66bp) -- (331bp,66bp) -- cycle;
  \draw (320bp,76bp) node {$q_8$};
\end{scope}
\begin{scope}
  \definecolor{strokecol}{rgb}{0.0,0.0,0.0};
  \pgfsetstrokecolor{strokecol}
  \draw (378bp,86bp) -- (356bp,86bp) -- (356bp,66bp) -- (378bp,66bp) -- cycle;
  \draw (367bp,76bp) node {$q_9$};
\end{scope}
\begin{scope}
  \definecolor{strokecol}{rgb}{0.0,0.0,0.0};
  \pgfsetstrokecolor{strokecol}
  \draw (100bp,20bp) -- (78bp,20bp) -- (78bp,0bp) -- (100bp,0bp) -- cycle;
  \draw (89bp,10bp) node {$q_3$};
\end{scope}
\begin{scope}
  \definecolor{strokecol}{rgb}{0.0,0.0,0.0};
  \pgfsetstrokecolor{strokecol}
  \draw [rounded corners] (273bp,144bp) ellipse (20bp and 12bp);
  \draw (273bp,144bp) node {$Q_{3,1}$};
\end{scope}
\begin{scope}
  \definecolor{strokecol}{rgb}{0.0,0.0,0.0};
  \pgfsetstrokecolor{strokecol}
  \draw [rounded corners] (320bp,144bp) ellipse (20bp and 12bp);
  \draw (320bp,144bp) node {$Q_{3,2}$};
\end{scope}
\begin{scope}
  \definecolor{strokecol}{rgb}{0.0,0.0,0.0};
  \pgfsetstrokecolor{strokecol}
  \draw [rounded corners] (367bp,144bp) ellipse (20bp and 12bp);
  \draw (367bp,144bp) node {$Q_{3,3}$};
\end{scope}
\begin{scope}
  \definecolor{strokecol}{rgb}{0.0,0.0,0.0};
  \pgfsetstrokecolor{strokecol}
  \draw [rounded corners] (226bp,144bp) ellipse (20bp and 12bp);
  \draw (226bp,144bp) node {$Q_{2,2}$};
\end{scope}
\begin{scope}
  \definecolor{strokecol}{rgb}{0.0,0.0,0.0};
  \pgfsetstrokecolor{strokecol}
  \draw [rounded corners] (179bp,144bp) ellipse (20bp and 12bp);
  \draw (179bp,144bp) node {$Q_{2,1}$};
\end{scope}
\begin{scope}
  \definecolor{strokecol}{rgb}{0.0,0.0,0.0};
  \pgfsetstrokecolor{strokecol}
  \draw [rounded corners] (201bp,274bp) ellipse (14bp and 12bp);
  \draw (201bp,274bp) node {$Q_\varnothing$};
\end{scope}
\begin{scope}
  \definecolor{strokecol}{rgb}{0.0,0.0,0.0};
  \pgfsetstrokecolor{strokecol}
  \draw [rounded corners] (114bp,210bp) ellipse (16bp and 12bp);
  \draw (114bp,210bp) node {$Q_1$};
\end{scope}
\begin{scope}
  \definecolor{strokecol}{rgb}{0.0,0.0,0.0};
  \pgfsetstrokecolor{strokecol}
  \draw [rounded corners] (201bp,210bp) ellipse (16bp and 12bp);
  \draw (201bp,210bp) node {$Q_2$};
\end{scope}
\begin{scope}
  \definecolor{strokecol}{rgb}{0.0,0.0,0.0};
  \pgfsetstrokecolor{strokecol}
  \draw [rounded corners] (296bp,210bp) ellipse (16bp and 12bp);
  \draw (296bp,210bp) node {$Q_3$};
\end{scope}
\begin{scope}
  \definecolor{strokecol}{rgb}{0.0,0.0,0.0};
  \pgfsetstrokecolor{strokecol}
  \draw [rounded corners] (89bp,76bp) ellipse (23bp and 12bp);
  \draw (89bp,76bp) node {$Q_{1,2,1}$};
\end{scope}
\begin{scope}
  \definecolor{strokecol}{rgb}{0.0,0.0,0.0};
  \pgfsetstrokecolor{strokecol}
  \draw [rounded corners] (142bp,76bp) ellipse (23bp and 12bp);
  \draw (142bp,76bp) node {$Q_{1,2,2}$};
\end{scope}
\end{tikzpicture}}
\caption{An example for the hierarchical organization of the physical degrees of
freedom $q_f$ into logical modes $Q_z$.  The edges are labeled with the numbers
$N_f$ of primitive basis functions for DOF $f$ or the numbers $n_z$ of SPFs for
node $z$, respectively.}
\label{fig:mode-tree}
\end{figure}

The successive splitting of the modes into sub-modes, sub-sub-modes etc.
leads to a hierarchy of modes, which is best visualized as a tree structure (the \emph{ML-tree}),
as exemplified by \reffig{fig:mode-tree}. Then each (sub\ldots-)mode can be
labeled by its corresponding node in this tree structure. A node $z$ is
specified by the path through which it can be reached from the top node,
i.e. $z=(\kappa,\lambda,\ldots)$. To specify that the node $z'$ is the $\mu$-th
child node of $z$, one appends $\mu$ to this path, in short $z' = (z,\mu)$.
For consistency of notation, the top node
itself is specified by an empty path, $z=\topn$. Its corresponding mode
$Q_\topn$ is the \emph{total mode} containing all degrees of freedom, and
its sub-modes are the original modes $Q_1, \ldots, Q_p$. 
In general, a mode $Q_z$ either has $p_z$ sub-modes $Q_{z,1},\ldots,Q_{z,p_z}$,
or it contains $d_z$ degrees of freedom $q_{a_z},\ldots,q_{b_z}$ ($d_z=b_z-a_z+1$).
In the first case, $z$ is called an \emph{internal} node, while in the latter case,
$z$ is called a \emph{leaf} node and $Q_z$ is called a \emph{primitive} mode.
Leaf nodes may also be labeled by its DOFs, i.e. $z=\prim{a_z,\ldots,b_z}$.

To shorten the notation, observe that the SPFs $\varphi^{(z)}_i(Q_z)$
are elements of a Hilbert space $\mathcal{H}^{(z)}$ (usually chosen as the
space of square-integrable functions over $Q_z$). Hence by specifying the
superscript, it is clear to which Hilbert space $\varphi^{(z)}_i$ belongs, and on which
mode it depends. Then the total wavefunction can be concisely expressed
as a tensor product, i.e. \refeq{eq:totwf} becomes
\begin{align}
\psi(t) = \sum_{i_1=1}^{n_1} \cdots \sum_{i_p=1}^{n_p}
A_{i_1 \cdots i_p}(t) \, \varphi^{(1)}_{i_1}(t) \otimes \cdots \otimes \varphi^{(p)}_{i_p}(t)
\:,
\label{eq:totwf1}
\end{align}
and likewise \refeq{eq:l1spf} can be re-expressed as
\begin{align}
\varphi^{(\kappa)}_i (t) &=
    \sum_{j_1=1}^{n_{\kappa,1}} \cdots \sum_{j_{p_\kappa}=1}^{n_{\kappa,p_\kappa}}
    A^{(\kappa)}_{i;j_1 \cdots j_{p_\kappa}}\!(t)
    \nonumber\\&\qquad\times
    \varphi^{(\kappa,1)}_{j_1}(t) \otimes \cdots \otimes \varphi^{(\kappa,p_\kappa)}_{j_{p_\kappa}}(t)
\:.
\label{eq:l1spf1}
\end{align}
More generally, for internal nodes, the SPFs are expressed as
\begin{align}
\varphi^{(z)}_i (t) &=
    \sum_{j_1=1}^{n_{z,1}} \cdots \sum_{j_{p_z}=1}^{n_{z,p_z}}
    A^{(z)}_{i;j_1 \cdots j_{p_z}}\!(t)
    \nonumber\\&\qquad\times
    \varphi^{(z,1)}_{j_1}(t) \otimes \cdots \otimes \varphi^{(z,p_z)}_{j_{p_z}}(t)
\:,
\label{eq:spf-int}
\end{align}
while for leaf nodes, they are expressed in terms of primitive basis functions,
\begin{align}
\varphi^{(z)}_i (t) =
    \sum_{\alpha_1=1}^{N_{a_z}} \cdots \sum_{\alpha_{d_z}=1}^{N_{b_z}}
    A^{(z)}_{i;\alpha_1 \cdots \alpha_{d_z}}\!(t)\,
    \chi^\prim{a_z}_{\alpha_1} \otimes \cdots \otimes \chi^\prim{b_z}_{\alpha_{d_z}}
\:.
\label{eq:spf-prim}
\end{align}

To further simplify notation, the time-dependence of the SPFs $\varphi^{(z)}_i$
and of the expansion coefficients $A^{(z)}_{i;\cdots}$ should by now be
clear and will no longer be mentioned unless necessary.
Additionally, by defining a multi-index $J^z$ for node $z$ as 
\begin{align}
J^z = \begin{cases}
(j_1, \ldots, j_{d_z})
\:\:;\: j_f = 1 \ldots N_{a_z+f-1}
&\text{if $z$ is a leaf}
\\
(j_1, \ldots, j_{p_z})
\:\:;\: j_\kappa = 1 \ldots n_{z,\kappa}
&\text{\hskip-1em if $z$ is internal}
\end{cases}
\end{align}
and by introducing the
\emph{configurations} $\Phi^{(z)}_{J^z}$ as
\begin{align}
\Phi^{(z)}_{J^z} =
\begin{cases}
\chi^\prim{a_z}_{j_1} \otimes \cdots \otimes \chi^\prim{b_z}_{j_{d_z}}
&\text{if $z$ is a leaf}
\\[1ex]
\varphi^{(z,1)}_{j_1} \otimes \cdots \otimes \varphi^{(z,p_z)}_{j_{p_z}}
&\text{if $z$ is internal}
\end{cases}
\label{eq:confs}
\end{align}
all SPFs can consistently be expanded as
\begin{align}
\varphi^{(z)}_i = \sum_{J^z} A^{(z)}_{i;J^z} \Phi^{(z)}_{J^z}
\quad.
\label{eq:spf-all}
\end{align}
\refeq{eq:spf-all} can even include the total wavefunction $\psi$ if one
uses $p_\topn=p$, $n_\topn=1$, and sets $\varphi^\topn_1=\psi$. That
is, the total wavefunction can be seen as a single SPF depending on the
total mode. For simplicity, one may omit the index $i=1$, as it is implied by
the top node, and use $\varphi^\topn$ for the total wavefunction:
\begin{align}
\varphi^\topn = \sum_{J^\topn} A^\topn_{J^\topn} \Phi^\topn_{J^\topn}
\label{eq:totwf2}
\end{align}

It is often useful to distinguish the part of the total wavefunction that
depends on a certain mode $Q_z$ from the part which depends on all
other coordinates, $Q_\topn \wo Q_z$.  The part that depends on $Q_z$ can be expressed
by the SPFs of node $z$, while the complementary
part defines the \emph{single-hole functions} (SHFs) $\Psi^{(z)}_l$:
\begin{align}
\label{eq:totwf-shfspf}
\varphi^\topn(Q_\topn) = \sum_{l=1}^{n_z}
\Psi^{(z)}_l(Q_\topn \wo Q_z)\,\varphi^{(z)}_l(Q_z)
\end{align}
That is, the SHFs are given by projecting out one particular
SPF from the total wavefunction, namely
\begin{align}
\label{eq:shf-def}
\Psi^{(z)}_l = \langle \varphi^{(z)}_l | \varphi^\topn \rangle
\quad.
\end{align}
For the top node, this leads to one trivial SHF, namely 
$\Psi^\topn = \langle \varphi^\topn | \varphi^\topn \rangle = 1$,
which does not depend on any coordinates.  For other nodes, there is
a useful recursion relation for expressing the SHFs. Any node $z'$ other
than the top node is a child of some node $z$, i.e. $z'=(z,\kappa)$ if it
is the $\kappa$-th child. The SHFs of such a node $z'$ can be written as
\begin{align}
\Psi^{(z,\kappa)}_l
& = \langle \varphi^{(z,\kappa)}_l | \varphi^\topn \rangle
    = \Bigl\langle \varphi^{(z,\kappa)}_l \Bigr|\Bigr. \sum_{m=1}^{n_z} \Psi^{(z)}_m \otimes \varphi^{(z)}_m \Bigr\rangle
\nonumber\\
& = \Bigl\langle \varphi^{(z,\kappa)}_l \Bigl|\Bigr. \sum_{m=1}^{n_z} \Psi^{(z)}_m \otimes
   \sum_{J^z} A^{(z)}_{m;J^z} \bigotimes_{\lambda=1}^{p_z} \varphi^{(z,\lambda)}_{j_\lambda} \Bigr\rangle
\nonumber\\
&= \sum_{m=1}^{n_z} \sum_{\multidxskip{J}{z}{\kappa}}
   A^{(z)}_{m;\multidxrepl{J}{z}{\kappa}{l}} \Psi^{(z)}_m \otimes
   \bigotimes_{\lambda \neq \kappa} \varphi^{(z,\lambda)}_{j_\lambda}
\label{eq:shf-rec}
\end{align}
where $\multidxskip{J}{z}{\kappa}$ and  $\multidxrepl{J}{z}{\kappa}{l}$ denote
multi-indices with a skipped or replaced index,
\begin{align}
\multidxskip{J}{z}{\kappa} &= (j_1, \ldots, j_{\kappa-1}, \phantom{l,}\, j_{\kappa+1}, \ldots, j_{p_z}) \\
\multidxrepl{J}{z}{\kappa}{l} &= (j_1, \ldots, j_{\kappa-1}, l, j_{\kappa+1}, \ldots, j_{p_z})
\quad.
\end{align}
\refeq{eq:shf-rec} means that the SHFs of a node $(z,\kappa)$ can be expressed
in terms of the SHFs and $A$-coefficients of its parent node $z$ plus the SPFs of the
sibling nodes $(z,\lambda)$, $\lambda \neq \kappa$.
Specifically, the $A$-coefficients of node $(z,\kappa)$ and its children are \emph{not} needed
to express its SHFs.

\subsection{Equations of motion}
\label{sec:eom}

The time-evolution of the total wavefunction under the action
of a Hamiltonian operator $\hat{H}$ can be obtained by inserting the
recursive expansion \refeqq{eq:spf-all}{eq:confs}  into the Dirac-Frenkel\cite{fre34}
variational principle,
\begin{align}
\langle \delta \varphi^\topn
| \hat{H} - i\frac{\partial}{\partial t} |
\varphi^\topn \rangle = 0
\quad.
\end{align}
The quantities being varied are all the SPFs of all nodes throughout
the ML-tree, which means that all coefficients $A^{(z)}_{l;J^z}$ are
being varied.  In order to maintain the orthonormality of the SPFs,
one may impose the constraints
\begin{align}
i\langle \varphi^{(z)}_l | \dot{\varphi}^{(z)}_m \rangle
&= \langle \varphi^{(z)}_l | \hat{g}^{(z)} | \varphi^{(z)}_m \rangle
\nonumber\\
&\quad \forall z \neq \topn \quad \forall\, l,m=1 \ldots n_z
\end{align}
where $\hat{g}^{(z)}$ is a Hermitian (but otherwise arbitrary) operator
acting on $Q_z$. In practice, $\hat{g}^{(z)}=0$ is used, and the further
discussion will be restricted to this case.

The variational procedure leads to equations of motion (EOMs) for the
$A$-coefficients. For the top node, one obtains
\begin{align}
\label{eq:eom-top}
i \dot{A}^\topn_{J^\topn} = \sum_{K^\topn} \,\langle \Phi^\topn_{J^\topn} | \hat{H} | \Phi^\topn_{K^\topn} \rangle \, A^\topn_{K^\topn}
\quad,
\end{align}
and for all other nodes
\begin{align}
\label{eq:eom-int}
i \dot{A}^{(z)}_{l;J^z} &=
    \sum_{K^z} \,\Bigl\langle \Phi^{(z)}_{J^z} \Bigl|
    (1-\hat{P}^{(z)})
    \nonumber\\[-1ex]&\qquad\times
    \sum_{i=1}^{n_z} (\rho^{(z)})^{-1}_{li} \sum_{m=1}^{n_z}
    \langle \hat{H} \rangle^{(z)}_{im} \Bigr| \Phi^{(z)}_{K^z} \Bigr\rangle \,A^{(z)}_{K^z}
\end{align}
where $\hat{P}^{(z)}$ is a projector onto the space spanned by the SPFs of node $z$,
\begin{align}
\hat{P}^{(z)} &=
    \sum_{l=1}^{n_z} |\varphi^{(z)}_l \rangle \langle \varphi^{(z)}_l |
    \\
    &= \sum_{l=1}^{n_z} \sum_{J^z} \sum_{K^z} A^{(z)}_{l;J^z} A^{(z)*}_{l;K^z} | \Phi^{(z)}_{J^z} \rangle \langle \Phi^{(z)}_{K^z} |
\quad,
\end{align}
$(\rho^{(z)})^{-1}$ is the inverse of the \emph{density matrix}  $\rho^{(z)}$ whose elements are given by
the overlaps of the SHFs of node $z$,
\begin{align}
\rho^{(z)}_{lm} = \langle \Psi^{(z)}_l | \Psi^{(z)}_m \rangle
\quad,
\end{align}
and $\langle \hat{H} \rangle^{(z)}$ is a matrix of \emph{mean-field operators} defined by
\begin{align}
\label{eq:mfop-def}
\langle\hat{H}\rangle^{(z)}_{lm} = \langle \Psi^{(z)}_l | \hat{H} | \Psi^{(z)}_m \rangle
\quad.
\end{align}
Note that each $\langle\hat{H}\rangle^{(z)}_{lm}$ is an operator on the coordinate $Q_z$,
as \refeq{eq:mfop-def} only integrates over the coordinates $Q_\topn \wo Q_z$.

\subsection{Hamiltonian in Sum-of-Products Form}
\label{sec:mlmctdh-sop}

In this and the following sections, the expressions involving leaf nodes will be
restricted to the case where the primitive modes contain only a single DOF. While
the more general case is very relevant in practice, this restriction simplifies the
notation, and the generalization of the expressions given here to primitive modes
with more than one DOF is straightforward.

\begingroup
\belowdisplayskip=12pt plus 3pt minus 7pt
\abovedisplayskip=12pt plus 3pt minus 7pt
In order for ML-MCTDH to be efficient, one needs a fast way to evaluate the right-hand side
of the equations of motion (\ref{eq:eom-top},\ref{eq:eom-int}).  In these, the most expensive part
is the computation of the terms
\begin{align}
\bigl\langle \Phi^{(z)}_{J^z} \bigr| \langle \hat{H} \rangle^{(z)}_{lm} \bigl| \Phi^{(z)}_{K^z} \bigr\rangle
= \bigl\langle \Phi^{(z)}_{J^z} \otimes \Psi^{(z)}_l \bigr| \hat{H} \bigl| \Phi^{(z)}_{K^z} \otimes \Psi^{(z)}_m \bigr\rangle
\quad.
\label{eq:full-terms}
\end{align}
Each of these terms constitutes (formally) a $d$-dimensional integration.
This \emph{quadrature problem} can be solved rather efficiently if the Hamiltonian
has the form of a sum of products of one-dimensional operators,
\begin{align}
\hat{H} = \sum_{r=1}^s c_r \hat{H}_r
\quad\text{with}\quad
\hat{H}_r = \bigotimes_{f=1}^d \hat{h}^\prim{f}_r
\quad.
\label{eq:sumprod}
\end{align}
For each node $z$, each $\hat{H}_r$ can be factored into an operator $\hat{h}^{(z)}_r$ acting on $Q_z$
and an operator $\hat{\mathcal{H}}^{(z)}_r$ acting on $Q_\topn \wo Q_z$:
\begin{align}
\hat{H}_r = \hat{h}^{(z)}_r \otimes \hat{\mathcal{H}}^{(z)}_r =
\bigotimes_{f \in Q_z } \hat{h}^\prim{f}_r
\; \otimes \;
\bigotimes_{f \notin Q_z } \hat{h}^\prim{f}_r
\end{align}
As Manthe has shown\cite{man08:164116}, 
this structure of the Hamiltonian permits a recursive scheme for evaluating 
the terms from \refeq{eq:full-terms}. They can be expressed as
\begin{align}
&\bigl\langle \Phi^{(z)}_{J^z} \bigr| \langle \hat{H} \rangle^{(z)}_{lm} \bigl| \Phi^{(z)}_{K^z} \bigr\rangle
\nonumber\\
&= \begin{cases} \displaystyle
    \sum_{r=1}^s c_r \, \mathfrak{H}^{(z)}_{r,lm} \,
    \bigl\langle \chi^\prim{f}_{J^z} \bigl| \hat{h}^\prim{f}_r \bigr| \chi^\prim{f}_{K^z} \bigr\rangle
    & \text{if $z$ is a leaf, $z=\prim{f}$}
    \\[1em] \displaystyle
    \sum_{r=1}^s c_r \, \mathfrak{H}^{(z)}_{r,lm}
    \prod_{\kappa=1}^{p_z} \mathfrak{h}^{(z,\kappa)}_{r,j_\kappa k_\kappa} & \text{otherwise}
    \end{cases}
\label{eq:full-terms-rec}
\end{align}
where the $\mathfrak{H}$- and $\mathfrak{h}$-terms
are defined by
\begin{align}
\mathfrak{h}^{(z)}_{r,jk}
& = \langle \varphi^{(z)}_j | \hat{h}^{(z)}_r | \varphi^{(z)}_k \rangle
\\
\mathfrak{H}^{(z)}_{r,lm}
& = \langle \Psi^{(z)}_l | \hat{\mathcal{H}}^{(z)}_r | \Psi^{(z)}_m \rangle
\end{align}
and which can be evaluated recursively:
\begin{align}
\mathfrak{h}^{(z)}_{r,jk}
   &= \begin{cases} \displaystyle
        \sum_{\alpha=1}^{N_f} \sum_{\beta=1}^{N_f} A^{(z)*}_{j;\alpha} A^{(z)}_{k;\beta}
        \, \bigl\langle \chi^\prim{f}_{\alpha} \bigl| \hat{h}^\prim{f}_r \bigr| \chi^\prim{f}_{\beta} \bigr\rangle
        & \text{if $z=\prim{f}$}
        \\ \displaystyle
        \sum_{I^z} \sum_{L^z} A^{(z)*}_{j;I^z} A^{(z)}_{k;L^z}
        \prod_{\kappa=1}^{p_z} \mathfrak{h}^{(z,\kappa)}_{r,i_\kappa l_\kappa} & \text{otherwise}
        \end{cases}
\label{eq:hterms}
\\
\mathfrak{H}^{\topn}_{r,11} &= 1
\\
\mathfrak{H}^{(z,\kappa)}_{r,lm}
   &= \sum_{a=1}^{n_z} \sum_{b=1}^{n_z} \mathfrak{H}^{(z)}_{r,ab}
        \sum_{\multidxskip{J}{z}{\kappa}} \sum_{\multidxskip{K}{z}{\kappa}}
        A^{(z)*}_{a;\multidxrepl{J}{z}{\kappa}{l}} A^{(z)}_{b;\multidxrepl{K}{z}{\kappa}{m}}
        \prod_{\lambda\neq\kappa} \mathfrak{h}^{(z,\lambda)}_{r,j_\lambda k_\lambda}
\label{eq:mfterms}
\end{align}
\refeq{eq:hterms} expresses the $\mathfrak{h}$-terms for node $z$ in terms of the
$\mathfrak{h}$-terms of its child nodes $(z,\kappa)$, which suggests a recursive
``bottom-up'' evaluation order, starting at the bottom layer where
the $\mathfrak{h}$-terms depend on the matrix elements
of the one-dimensional operators $\hat{h}^\prim{f}_r$ in the primitive bases.
On the other hand, \refeq{eq:mfterms} expresses the $\mathfrak{H}$-terms for a
node $(z,\kappa)$ in terms of the $\mathfrak{H}$-terms of its parent node $z$ and
the $\mathfrak{h}$-terms of its sibling nodes $(z,\lambda), \lambda\neq\kappa$,
leading to a ``top-down'' recursive evaluation, starting from
the top node where $\mathfrak{H}^\topn_r = 1$.
\endgroup

\begin{widetext}
For a Hamiltonian in form (\ref{eq:sumprod}), using \refeq{eq:full-terms-rec} yields
the following explicit EOMs for the $A$-coefficients: For the top node
\begin{align}
i \dot{A}^\topn_{J^\topn}
=& \sum_{r=1}^s c_r \sum_{K^\topn}  A^\topn_{K^\topn} \prod_{\kappa=1}^{p_\topn} \mathfrak{h}^{(\kappa)}_{r,j_\kappa k_\kappa}
\:,
\label{eq:eompf-top}
\\
\intertext{for any other internal node $z$}
i \dot{A}^{(z)}_{l;J^z}
=& \sum_{K^z} \Bigl( \delta_{J^z K^z} - \sum_{a=1}^{n_z} A^{(z)}_{a;J^z} A^{(z)*}_{a;K^z} \Bigr)
    \sum_{l'=1}^{n_z} (\rho^{(z)})^{-1}_{ll'}
    \sum_{r=1}^s c_r \sum_{m=1}^{n_z} \mathfrak{H}^{(z)}_{r,l'm} \sum_{L^z}  A^{(z)}_{m;L^z}
    \prod_{\kappa=1}^{p_z} \mathfrak{h}^{(z,\kappa)}_{r,k_\kappa l_\kappa}
\:,
\label{eq:eompf-int}
\\
\intertext{and for a leaf node $z=\prim{f}$}
i \dot{A}^{(z)}_{l;\alpha}
=& \sum_{\beta=1}^{N_f} \Bigl( \delta_{\alpha \beta} - \sum_{a=1}^{n_z} A^{(z)}_{a;\alpha} A^{(z)*}_{a;\beta} \Bigr)
    \sum_{l'=1}^{n_z} (\rho^{(z)})^{-1}_{ll'}
    \sum_{r=1}^s c_r \sum_{m=1}^{n_z} \mathfrak{H}^{(z)}_{r,l'm} \sum_{\gamma=1}^{N_f}  A^{(z)}_{m;\gamma}
    \bigl\langle \chi^\prim{f}_{\beta} \bigl| \hat{h}^\prim{f}_r \bigr| \chi^\prim{f}_{\gamma} \bigr\rangle
\:.
\label{eq:eompf-leaf}
\end{align}
\end{widetext}

It is evident that the numerical effort for evaluating Eqs.\,(\ref{eq:hterms}--\ref{eq:eompf-leaf})
scales linearly with $s$, the number of terms in the Hamiltonian expansion. A detailed
analysis of this numerical effort will be given in \refapp{app:numeff-pf}.

\section{ML-MCTDH with multi-layer operators}
\label{sec:mlop+mlmctdh}

The sum-of-products operator form, \refeq{eq:sumprod}, has traditionally been
used for MCTDH calculations, and as the previous section has shown, it is also
well-suited for ML-MCTDH calculations. However, in comparison to MCTDH, the
ML-MCTDH wavefunction possesses an additional \emph{hierarchical} structure, 
and it is worthwile exploring whether a similar structure for the Hamiltonian
operator leads to additional benefits. In this section, the theory for such
\emph{multi-layer operators} and their application within ML-MCTDH is developed.

\subsection{Multi-layer operators}
\label{sec:mlop}

Consider an operator that has, like the wavefunction, a hierarchical
structure. At the top level, such an operator reads analogous to \refeq{eq:totwf2}
\begin{align}
\hat{U}^\topn
&= \sum_{c_1=1}^{m_1} \cdots \sum_{c_{p_\topn}=1}^{m_{p_\topn}} V^\topn_{c_1 \cdots c_{p_\topn}}
    \hat{U}^{(1)}_{c_1} \otimes \cdots \otimes \hat{U}^{(p_\topn)}_{c_{p_\topn}}
\nonumber\\
&= \sum_{C^\topn} V^\topn_{C^\topn} \bigotimes_{\kappa=1}^{p_\topn} \hat{U}^{(\kappa)}_{c_\kappa}
\label{def:hierop}
\end{align}
where $\hat{U}^\topn$ operates on all degrees of freedom $Q_\topn$, while the
$\hat{U}^{(\kappa)}_{c_\kappa}$ operate only on $Q_\kappa$. The first-layer operators
$\hat{U}^{(\kappa)}_{c}$ are then expanded in a similar manner as sums of products
of second-layer operators, and so on.  This recursive expansion of $\hat{U}^\topn$ stops
at the bottom layer, where the leaf operators  act directly on primitive modes.
Reusing the terminology from ML-MCTDH, the operators $\hat{U}^{(z)}_b$ shall be called
\emph{single-particle operators} (SPOs). Their expansion reads
\begin{align}
\hat{U}^{(z)}_b &= \begin{cases}
\hat{U}^\prim{f}_b
& \text{if $z=\prim{f}$ is a leaf}
\\ \displaystyle
\sum_{C^z} V^{(z)}_{b;C^z} \bigotimes_{\kappa=1}^{p_z} \hat{U}^{(z,\kappa)}_{c_\kappa}
& \text{otherwise}
\end{cases}
\label{def:hierop2}
\\
&( b=1 \ldots m_z ) \nonumber
\end{align}
where $C^z$ is again a multi-index, i.e. $C^z=(c_1,\ldots,c_{p_z})$ with $c_\kappa = 1 \ldots m_{z,\kappa}$.
The leaf operators $\hat{U}^\prim{f}_b$ and the \emph{transfer tensors} $V^{(z)}_{b;C^z}$
together determine the full operator $\hat{U}^\topn$, in the same manner as the
primitive basis functions $\chi^\prim{f}_\alpha$ and the the $A$-coefficients determine
the total wavefunction.
An operator that has the hierarchical structure (\ref{def:hierop},\ref{def:hierop2}) shall be
called a \emph{multi-layer operator}.

Like for the multi-layer wavefunction, it can be useful to decompose the multi-layer
operator $\hat{U}^\topn$ into a part acting on $Q_z$ and a
part acting on all other degrees of freedom:
\begin{align}
\hat{U}^\topn = \sum_{b=1}^{m_z} \hat{W}^{(z)}_b \otimes \hat{U}^{(z)}_b
\label{eq:sposho}
\end{align}
This introduces the \emph{single-hole operators} (SHOs) $\hat{W}^{(z)}_b$.
For example, if $z$ is a child of the top node, i.e. $z=(\kappa)$, the explicit
form of $\hat{W}^{(\kappa)}_b$ can be seen directly from \refeq{def:hierop}:
\begin{align}
\hat{W}^{(\kappa)}_b 
= \sum_{\multidxskip{C}{\topn}{\kappa}} V^\topn_{\multidxrepl{C}{\topn}{\kappa}{b}}
    \bigotimes_{\lambda \neq \kappa} \hat{U}^{(\lambda)}_{c_\lambda}
\end{align}
In general an explicit form for the SHOs will be cumbersome to write down,
but like for the SHFs one can obtain a recursive form of the SHOs by decomposing
$\hat{U}^\topn$ into SHOs and SPOs for a node $(z,\kappa)$ and for its parent $z$:
\begin{align}
\hat{U}^\topn
&= \sum_{a=1}^{m_{z,\kappa}} \hat{W}^{(z,\kappa)}_a \otimes \hat{U}^{(z,\kappa)}_a
\label{eq:recsho1}
\\
&= \sum_{b=1}^{m_z} \hat{W}^{(z)}_b \otimes \hat{U}^{(z)}_b
   = \sum_{b=1}^{m_z} \hat{W}^{(z)}_b \otimes \sum_{C^z} V^{(z)}_{b;C^z}
    \bigotimes_{\lambda=1}^{p_z} \hat{U}^{(z,\lambda)}_{c_\lambda}
\nonumber\\
&= \sum_{a=1}^{m_{z,\kappa}} \left( \sum_{b=1}^{m_z} \sum_{\multidxskip{C}{z}{\kappa}}
    V^{(z)}_{b;\multidxrepl{C}{z}{\kappa}{a}} \hat{W}^{(z)}_b \otimes
    \bigotimes_{\lambda\neq\kappa}  \hat{U}^{(z,\lambda)}_{c_\lambda}
    \right) \otimes \hat{U}^{(z,\kappa)}_a
\label{eq:recsho2}
\end{align}
In the last step, $c_\kappa$ was simply renamed to $a$.
Comparing \refeq{eq:recsho1} and \refeq{eq:recsho2} yields
an expression for an SHO in terms of the parent SHOs and the
sibling SPOs:
\begin{align}
\hat{W}^{(z,\kappa)}_a = \sum_{b=1}^{m_z} \sum_{\multidxskip{C}{z}{\kappa}}
    V^{(z)}_{b;\multidxrepl{C}{z}{\kappa}{a}} \hat{W}^{(z)}_b \otimes
    \bigotimes_{\lambda\neq\kappa}  \hat{U}^{(z,\lambda)}_{c_\lambda}
\label{eq:recsho}
\end{align}
This is the operator analogue to \refeq{eq:shf-rec}.
For completeness, note that for the top node,
$m_\topn=1$, $\hat{W}^\topn_1=1$, and $\hat{U}^\topn_1=\hat{U}^\topn$
hold trivially.

\subsection{EOMs with multi-layer operators}
\label{sec:mlop-eom}

The ML-MCTDH equations of motion in the form (\ref{eq:eom-top},\ref{eq:eom-int})
are independent of the form of the Hamiltonian. As stated previously, the
evaluation of the terms from \refeq{eq:full-terms} is central to their efficiency.
Concerning the use of multi-layer operators, note that in general the Hamiltonian
$\hat{H}$ may be a sum of a sum-of-products operator and several multi-layer operators.
However, as the EOMs are linear in $\hat{H}$, it is sufficient to consider the contributions
of one multi-layer operator $\hat{U}^\topn$ to the terms (\ref{eq:full-terms}).

\begingroup
\belowdisplayskip=12pt plus 3pt minus 7pt
\abovedisplayskip=12pt plus 3pt minus 7pt
The following assumes that the multi-layer operator $\hat{U}^\topn$ uses
exactly the same tree structure as the ML-MCTDH wavefunction $\varphi^\topn$,
though the number of SPFs ($n_z$) and SPOs ($m_z$) per node may differ.
Using the decomposition from \refeq{eq:sposho}, one obtains the following
contribution to (\ref{eq:full-terms}):
\begin{align}
&\bigl\langle \Phi^{(z)}_{I^z} \bigl| \langle \hat{U}^\topn \rangle^{(z)}_{jk} \bigr| \Phi^{(z)}_{L^z} \bigr\rangle
\nonumber\\
&\quad= \Bigl\langle \Phi^{(z)}_{I^z} \otimes \Psi^{(z)}_j \,\Bigl|\,
    \sum_{b=1}^{m_z} \hat{W}^{(z)}_b \otimes \hat{U}^{(z)}_b
    \,\Bigr|\, \Phi^{(z)}_{L^z} \otimes \Psi^{(z)}_k \Bigr\rangle
\nonumber\\
&\quad= \sum_{b=1}^{m_z} \langle \Psi^{(z)}_j | \hat{W}^{(z)}_b | \Psi^{(z)}_k \rangle \,
    \langle \Phi^{(z)}_{I^z} | \hat{U}^{(z)}_b | \Phi^{(z)}_{L^z} \rangle
\end{align}
The second factor of this expression reads, for $z=\prim{f}$,
\begin{align}
\langle \Phi^{(z)}_{\alpha} | \hat{U}^{(z)}_b | \Phi^{(z)}_{\beta} \rangle
= \langle \chi^\prim{f}_\alpha | \hat{U}^\prim{f}_b | \chi^\prim{f}_\beta \rangle
=: \mathsf{U}^\prim{f}_{b,\alpha\beta}
\quad,
\end{align}
while for non-leaf $z$ it can be further decomposed:
\begin{align}
&\langle \Phi^{(z)}_{I^z} | \hat{U}^{(z)}_b | \Phi^{(z)}_{L^z} \rangle
\nonumber\\
&\quad= \Bigl\langle \bigotimes_{\kappa=1}^{p_z} \varphi^{(z,\kappa)}_{i_\kappa} \Bigl|
     \sum_{C^z} V^{(z)}_{b;C^z} \bigotimes_{\kappa=1}^{p_z} \hat{U}^{(z,\kappa)}_{c_\kappa} \Bigr|
     \bigotimes_{\kappa=1}^{p_z} \varphi^{(z,\kappa)}_{l_\kappa} \Bigr\rangle
\nonumber\\
&\quad= \sum_{C^z} V^{(z)}_{b;C^z} \prod_{\kappa=1}^{p_z}
    \bigl\langle  \varphi^{(z,\kappa)}_{i_\kappa} \bigl| \hat{U}^{(z,\kappa)}_{c_\kappa} \bigr| \varphi^{(z,\kappa)}_{l_\kappa} \bigr\rangle
\end{align}
Using the abbreviations
\begin{align}
\mathfrak{W}^{(z)}_{b,jk} &:= \langle \Psi^{(z)}_j | \hat{W}^{(z)}_b | \Psi^{(z)}_k \rangle
\\
\mathfrak{U}^{(z)}_{c,jk} &:= \langle  \varphi^{(z)}_j | \hat{U}^{(z)}_c | \varphi^{(z)}_k \rangle
\end{align}
one arrives at
\begin{align}
&\bigl\langle \Phi^{(z)}_{I^z} \bigl| \langle \hat{U}^\topn \rangle^z_{jk} \bigr| \Phi^{(z)}_{L^z} \bigr\rangle
\nonumber\\
&\quad=
  \sum_{b=1}^{m_z} \mathfrak{W}^{(z)}_{b,jk}
    \begin{cases}
        \displaystyle
        \mathsf{U}^\prim{f}_{b,I^z L^z}
        & \text{if $z=\prim{f}$}
        \\
        \displaystyle
        \sum_{C^z} V^{(z)}_{b;C^z} \prod_{\kappa=1}^{p_z} \mathfrak{U}^{(z,\kappa)}_{c_\kappa, i_\kappa l_\kappa}
        & \text{otherwise}
    \end{cases}
\:.
\label{eq:full-terms-mlpf}
\end{align}
Like in the sum-of-products case, it is possible to compute the quantities
$\mathfrak{W}^{(z)}_{b,jk}$ and $\mathfrak{U}^{(z)}_{c,jk}$ recursively.
For the $\mathfrak{U}$-terms at a non-leaf node $z$, express the SPFs
and SPOs in terms of those of its children:
\begin{align}
\mathfrak{U}^{(z)}_{c,jk}
&= \langle \varphi^{(z)}_j | \hat{U}^{(z)}_c | \varphi^{(z)}_k \rangle
\nonumber\\
&= \Bigl\langle \sum_{I^z} A^{(z)}_{j;I^z} \bigotimes_{\kappa=1}^{p_z} \varphi^{(z,\kappa)}_{i_\kappa}
   \,\Bigl|\,
   \sum_{B^z} V^{(z)}_{c;B^z} \bigotimes_{\kappa=1}^{p_z} \hat{U}^{(z,\kappa)}_{b_\kappa}
\nonumber\\
& \mkern28mu \Bigr|
   \sum_{L^z} A^{(z)}_{k;L^z} \bigotimes_{\kappa=1}^{p_z} \varphi^{(z,\kappa)}_{l_\kappa} \Bigr\rangle
\nonumber\\
&= \sum_{B^z} \sum_{I^z} \sum_{L^z} V^{(z)}_{c;B^z} A^{(z)*}_{j;I^z} A^{(z)}_{k;L^z}
   \prod_{\kappa=1}^{p_z} \langle  \varphi^{(z,\kappa)}_{i_\kappa} |
   \hat{U}^{(z,\kappa)}_{b_\kappa} | \varphi^{(z,\kappa)}_{l_\kappa} \rangle
\nonumber\\
&= \sum_{B^z} \sum_{I^z} \sum_{L^z} V^{(z)}_{c;B^z} A^{(z)*}_{j;I^z} A^{(z)}_{k;L^z}
   \prod_{\kappa=1}^{p_z} \mathfrak{U}^{(z,\kappa)}_{b_\kappa, i_\kappa l_\kappa}
\label{eq:uterms}
\end{align}
This shows that the $\mathfrak{U}$-terms at node $z$ can be computed from those of its children,
leading to a ``bottom-up'' recursive evaluation that starts from 
the $\mathfrak{U}$-terms at the leaf nodes, which are given by
\begin{align}
\mathfrak{U}^\prim{f}_{c,jk} 
&= \langle \varphi^\prim{f}_j | \hat{U}^\prim{f}_c | \varphi^\prim{f}_k \rangle
\nonumber\\
&= \Bigl\langle \sum_{\alpha=1}^{N_f} A^\prim{f}_{j;\alpha} \chi^\prim{f}_\alpha \,\Bigl|\,
    \hat{U}^\prim{f}_c \,\Bigr|\, \sum_{\beta=1}^{N_f} A^\prim{f}_{k;\beta} \chi^\prim{f}_\beta \Bigr\rangle
\nonumber\\
&= \sum_{\alpha=1}^{N_f}  \sum_{\beta=1}^{N_f} A^{\prim{f}\:*}_{j;\alpha} A^\prim{f}_{k;\beta}
\,\underbrace{\langle \chi^\prim{f}_\alpha | \hat{U}^\prim{f}_c | \chi^\prim{f}_\beta \rangle}%
_{\displaystyle \mathsf{U}^\prim{f}_{c,\alpha\beta}} \quad.
\label{eq:uterms-leaf}
\end{align}
Provided that the leaf operators $\hat{U}^\prim{f}_c$ are time-independent, their
matrix representation $\boldsymbol{\mathsf{U}}^\prim{f}_c$ in the primitive
basis needs to be computed only once in the beginning.
\endgroup

To derive a recursive expression for the $\mathfrak{W}$-terms,
one makes use of \refeq{eq:shf-rec} and \refeq{eq:recsho}. Then the
$\mathfrak{W}$-term for the $\kappa$-th child of node $z$ reads
\begin{align}
\mathfrak{W}^{(z,\kappa)}_{b,jk}
&= \langle \Psi^{(z,\kappa)}_j | \hat{W}^{(z,\kappa)}_b | \Psi^{(z,\kappa)}_k \rangle
\nonumber\\
&= \Bigl\langle
    \sum_{i=1}^{n_z} \sum_{\multidxskip{J}{z}{\kappa}} A^{(z)}_{i;\multidxrepl{J}{z}{\kappa}{j}}
        \Psi^{(z)}_i \otimes \bigotimes_{\lambda \neq \kappa} \varphi^{(z,\lambda)}_{j_\lambda}
\nonumber\\
& \mkern28mu \Bigr|\,
    \sum_{c=1}^{m_z} \sum_{\multidxskip{B}{z}{\kappa}} V^{(z)}_{c;\multidxrepl{B}{z}{\kappa}{b}}
        \hat{W}^{(z)}_c \otimes \bigotimes_{\lambda \neq \kappa} \hat{U}^{(z,\lambda)}_{b_\lambda}
\nonumber\\
& \mkern28mu \Bigl|\,
    \sum_{l=1}^{n_z} \sum_{\multidxskip{K}{z}{\kappa}} A^{(z)}_{l;\multidxrepl{K}{z}{\kappa}{k}}
        \Psi^{(z)}_l \otimes \bigotimes_{\lambda \neq \kappa} \varphi^{(z,\lambda)}_{k_\lambda}
    \Bigr\rangle
\nonumber\\
&= \sum_{c=1}^{m_z} \sum_{i=1}^{n_z} \sum_{l=1}^{n_z} \mathfrak{W}^{(z)}_{c,il} 
    \sum_{\multidxskip{B}{z}{\kappa}} \sum_{\multidxskip{J}{z}{\kappa}} \sum_{\multidxskip{K}{z}{\kappa}}
    V^{(z)}_{c;\multidxrepl{B}{z}{\kappa}{b}}
\nonumber\\ & \mkern28mu \times\:
    A^{(z)*}_{i;\multidxrepl{J}{z}{\kappa}{j}} A^{(z)}_{l;\multidxrepl{K}{z}{\kappa}{k}}
    \prod_{\lambda \neq \kappa} \mathfrak{U}^{(z,\lambda)}_{b_\lambda, j_\lambda k_\lambda}
\quad.
\label{eq:wterms}
\end{align}
Hence the $\mathfrak{W}$-terms of the parent and the $\mathfrak{U}$-terms
of the siblings are needed to compute the  $\mathfrak{W}$-terms of a node.
This leads to a ``top-down'' recursive evaluation that starts from
the $\mathfrak{W}$-term at the top node, which is simply
\begin{align}
\mathfrak{W}^\topn_{1,11}
= \langle \Psi^\topn_1 | \hat{W}^\topn_1 | \Psi^\topn_1 \rangle
= 1
\quad.
\end{align}

\begin{widetext}
\begingroup
\belowdisplayskip=9pt plus 0pt minus 4pt
\abovedisplayskip=9pt plus 0pt minus 4pt
Finally, from \refeq{eq:full-terms-mlpf} one gets the following contributions
of the hierarchical operator $\hat{U}^\topn$  to the RHS of the ML-MCTDH
equations of motion: For the top node
\begin{align}
i \dot{A}^{\topn}_{J^\topn} &= \cdots \: + \:
    \sum_{B^\topn} V^\topn_{B^\topn} \sum_{L^\topn} \prod_{\kappa=1}^{p_\topn}
    \mathfrak{U}^{(\kappa)}_{b_\kappa, j_\kappa l_\kappa} A^\topn_{L^\topn}
\quad,
\label{eq:eom-mlpf-top}
\\
\intertext{for any other internal node $z$}
i \dot{A}^{(z)}_{l;J^z} &= \cdots \: + \:
    \sum_{K^z} \Bigl( \delta_{J^z K^z}
        - \sum_{a=1}^{n_z} A^{(z)}_{a;J^z} A^{(z)*}_{a;K^z} \Bigr)
    \sum_{l'=1}^{n_z} (\rho^{(z)})^{-1}_{ll'}
    \sum_{c=1}^{m_z} \sum_{l''=1}^{n_z}
    \mathfrak{W}^{(z)}_{c,l'l''}
        \sum_{B^z} V^{(z)}_{c;B^z} \sum_{L^z}
        \prod_{\kappa=1}^{p_z} \mathfrak{U}^{(z,\kappa)}_{b_\kappa, k_\kappa l_\kappa}
        A^{(z)}_{l'';L^z}
\quad,
\label{eq:eom-mlpf-int}
\\
\intertext{and for a leaf node $z=\prim{f}$}
i \dot{A}^{(z)}_{l;\alpha} &= \cdots \: + \:
    \sum_{\beta=1}^{N_f} \Bigl( \delta_{\alpha \beta}
        - \sum_{a=1}^{n_z} A^{(z)}_{a;\alpha} A^{(z)*}_{a;\beta} \Bigr)
    \sum_{l'=1}^{n_z} (\rho^{(z)})^{-1}_{ll'} 
    \sum_{c=1}^{m_z} \sum_{l''=1}^{n_z}
    \mathfrak{W}^{(z)}_{c,l'l''}
        \sum_{\gamma=1}^{N_f} \mathsf{U}^\prim{f}_{c,\beta \gamma} A^{(z)}_{l'';\gamma}
\quad.
\label{eq:eom-mlpf-leaf}
\end{align}
\endgroup
\end{widetext}

Not surprisingly, the ML-MCTDH EOMs with multi-layer operators (including
the recursive expressions for the $\mathfrak{U}$- and $\mathfrak{W}$-terms)
are more complex than the corresponding expressions
(\ref{eq:hterms}--\ref{eq:eompf-leaf}) for sum-of-products
operators. However, in the sum-of-products case, one has to evaluate these
expressions for \emph{each} term in the sum, while the expressions for multi-layer
operators directly yield the result for the \emph{full} operator.  How this can lead to
performance gains will be the subject of the discussion in \refsec{sec:numeff}.

\section{Potfit and multi-layer Potfit}
\label{sec:pf+mlpf}

With the introduction of multi-layer operators, the question arises for which
parts of the Hamiltonian this new operator format may be beneficial.
As mentioned in the introduction, the kinetic energy operator (KEO) naturally
possesses a sum-of-products structure, provided that a suitable coordinate
system has been chosen. Moreover, the number of terms for the KEO is
usually not large, so that a conversion into a multi-layer operator is likely
not of much benefit, if any. Hence, this article will not explore the use of
multi-layer operators for the KEO.

In contrast, the potential energy part of the Hamiltonian is often described by
a potential energy surface (PES) which is given as a general function of all coordinates. 
The representation of such a PES in any format, whether sum-of-products or
multi-layer, necessitates a fitting procedure which introduces an approximation.
Hence one has to find a balance between
\begin{itemize}
\item
the \emph{quality} of this approximation,
\item
the computational cost for \emph{generating} the fit, and
\item
the resulting \emph{size} of the fit, which in turn influences the
computational cost for ML-MCTDH.
\end{itemize}
Several approaches for fitting a PES into sum-of-products form have
been mentioned in the introduction. Among these, the Potfit method offers
a superior control of accuracy, and it turns out that it can be rather
straight-forwardly generalized into a procedure for fitting a PES into a
multi-layer structure. After reviewing some essential properties of
Potfit, this section will discuss the resulting algorithm, \emph{multi-layer Potfit}.

\subsection{Review of Potfit}
\label{sec:potfit}

The Potfit algorithm introduced by Jäckle and Meyer\cite{jae96:7974}
can yield a PES representation in sum-of-products form, as long as
the dimensionality of the system is not too high (say, 6--8 physical degrees of freedom,
or 4--5 primitive modes).  The algorithm starts from evaluating the PES function
$V$ on a product grid, which results in a potential energy tensor of order $d$,
\begin{align}
V_{\alpha_1 \cdots \alpha_d} = V(q^\prim{1}_{\alpha_1}, \ldots, q^\prim{d}_{\alpha_d})
\quad; \quad\alpha_f = 1 \ldots N_f
\quad.
\end{align}
In practice, the grid points $q^\prim{f}_\alpha$ are closely related to the
primitive basis functions $\chi^\prim{f}_\alpha$, e.g. often a discrete
variable representation (DVR) is used, where the $\chi^\prim{f}_\alpha$ are localized such that
$\chi^\prim{f}_\alpha(q^\prim{f}_\beta) = \delta_{\alpha \beta}$. In such
a representation, the elements of the $V$-tensor can also be seen as
the diagonal matrix elements of the potential energy operator $\hat{V}$
in the primitive product basis (and its non-diagonal matrix elements vanish).
 
Next, this tensor is subjected to what has later become known in the mathematical
literature as a \emph{higher-order singular value decomposition} (HOSVD)\cite{lat00:1253}. For
each dimension $f$, one builds a \emph{potential density matrix} by contracting
the $V$-tensor with itself along all indices except the $f$-th, and then determines
the eigenvalues $\lambda^\prim{f}_i$ and eigenvectors $v^\prim{f}_i$ of this
symmetric and positive semidefinite matrix. Equivalently, one builds
a \emph{matricization} of the $V$-tensor by unfolding all indices except the $f$-th,
and then performs an SVD on the resulting matrix, which yields singular
values $\sigma^\prim{f}_i$ and singular vectors $v^\prim{f}_i$. Numerically, the SVD approach
is more stable, but requires (much) more computer memory.
If the eigenvalues/singular values are in descending order, then both approaches
are connected through $\lambda^\prim{f}_i = (\sigma^\prim{f}_i)^2$, and they
result in the same orthonormal vectors. In Potfit terminology,
these vectors are called \emph{natural potentials} and the $\lambda^\prim{f}_i$
are called \emph{natural weights}.  One can associate each natural potential
$v^\prim{f}_i$ with a one-dimensional local operator $\hat{v}^\prim{f}_i$ whose
matrix elements in the primitive basis are given by
$\langle \chi^\prim{f}_\alpha | \hat{v}^\prim{f}_i | \chi^\prim{f}_\beta \rangle
= v^\prim{f}_{i \alpha} \delta_{\alpha \beta}$.

By keeping, for each $f$, only the $m_f$ dominant natural potentials
(i.e. those with the largest natural weights),
one arrives at the following approximation for the $V$-tensor,
\begin{align}
V_{\alpha_1 \cdots \alpha_d} \approx \tilde{V} _{\alpha_1 \cdots \alpha_d} =
\sum_{i_1=1}^{m_1} \cdots \sum_{i_d=1}^{m_d} C_{i_1 \cdots i_d}\,
v^\prim{1}_{i_1 \alpha_1} \cdots v^\prim{d}_{i_d \alpha_d}
\quad,
\label{eq:potfit1-tensor}
\end{align}
or correspondingly for the potential operator,
\begin{align}
\hat{V} \approx \sum_{i_1=1}^{m_1} \cdots \sum_{i_d=1}^{m_d} C_{i_1 \cdots i_d}\,
\hat{v}^\prim{1}_{i_1} \otimes \cdots \otimes \hat{v}^\prim{d}_{i_d}
\quad.
\label{eq:potfit1}
\end{align}
\refeq{eq:potfit1-tensor} is commonly known as a \emph{Tucker representation}\cite{tuc63:122}.
The elements of the \emph{core tensor} $C$ are obtained by projecting
$V$ onto products of the natural potentials, i.e.
\begin{align}
C_{i_1 \cdots i_d} = \sum_{\alpha_1=1}^{N_1} \cdots \sum_{\alpha_d=1}^{N_d}
v^\prim{1}_{i_1 \alpha_1} \cdots v^\prim{d}_{i_d \alpha_d} V_{\alpha_1 \cdots \alpha_d}
\quad.
\label{eq:pfcore}
\end{align}
The error that is introduced by this approximation is bounded by\cite{jae98:3772,lat00:1253}
\begin{align}
\Vert V - \tilde{V} \Vert^2 \leq \sum_{f=1}^d \, \sum_{i_f > m_f} \, \lambda^\prim{f}_{i_f}
\label{eq:pf-error}
\end{align}
where $\Vert\cdot\Vert$ is the Frobenius norm.  A more physical error measure,
the root-mean-square (rms) error $\Delta_\text{rms}$, is related to this by
$\Delta_\text{rms} = \sqrt{\Vert V - \tilde{V} \Vert^2 / N_\text{grid}}$, where
$N_\text{grid} = \prod_{f=1}^d N_f$ is the total number of points in the product grid.

The expansion (\ref{eq:potfit1}) requires $\prod_{f=1}^d m_d$ terms in total.
This number may be reduced by actually carrying out one of the summations,
say over $i_1$ (and then one may as well choose $m_1=N_1$).  Using this
\emph{contraction} trick, one arrives at
\begin{align}
\hat{V} &\approx \sum_{i_2=1}^{m_2} \cdots \sum_{i_d=1}^{m_d}
\hat{D}^\prim{1}_{i_2 \cdots i_d} \otimes
\hat{v}^\prim{2}_{i_2} \otimes \cdots \otimes \hat{v}^\prim{d}_{i_d}
\label{eq:contrpf}
\\
\text{with}\quad&
\hat{D}^\prim{1}_{i_2 \cdots i_d} = \sum_{i_1=1}^{N_1} C_{i_1 \cdots i_d}\,\hat{v}^\prim{1}_{i_1}
\end{align}
Now the number of terms needed to represent $\hat{V}$ has been reduced
by a factor $m_1$.  In practice, one contracts over that index $f$ where $m_f$
would be largest.  Assuming that all primitive modes employ $m$ natural
potentials, the overall number of terms for representing the potential operator
(i.e. the number of summands in \refeq{eq:contrpf}) is $m^{d-1}$.

The quality of the Potfit approximation can be improved (or, given a target accuracy,
the number of necessary terms can be reduced) through an iterative procedure\cite{jae96:7974,jae98:3772}.
This is especially worthwhile if one introduces a \emph{relevant region}, i.e. a region of the product grid
where the potential representation should be of good quality (e.g. the region where the potential energy
lies below a certain threshold). For details, see \refcite{bec00:1}.

A limiting factor for Potfit is that one needs to initially evaluate and store the
potential on a full product grid. The resulting amount of data scales exponentially
like $N^d$, which makes the method infeasible for larger $d$. Recently, Peláez and Meyer
introduced the \emph{multigrid Potfit} method (MGPF)\cite{pel13:014108} which aims
to circumvent this limitation by employing two nested grids, a coarse one and a fine one.
In MGPF, the potential only needs to be evaluated on product grids that are fine in
one mode and coarse in all others. This drastically reduces the amount of data
that needs to be processed, and computations for up to about $12$ degrees of freedom
seem feasible with MGPF.  However, MGPF does \emph{not} reduce the number of terms
that are needed to represent the potential; this number still scales as $m^{d-1}$.

\subsection{The multi-layer generalization of Potfit}
\label{sec:mlpf}

At this point it may be helpful for the reader to recall how
the transition from MCTDH to ML-MCTDH is accomplished:
physical degrees of freedom are first combined into logical
modes, then these modes are combined into larger meta-modes,
and so on, resulting in a tree-like mode hierarchy  (cf. \reffig{fig:mode-tree}).
In short, the central idea is \emph{repeated mode combination}.
Combining this idea with the Potfit/HOSVD algorithm straight-forwardly
leads to the \emph{multi-layer Potfit} (MLPF) procedure. The
resulting method has been described previously in 
the mathematical literature by Grasedyck as 
\emph{hierarchical SVD with leaves-to-root truncation}
(Algorithm 2 in \refcite{gra10:2029}). A similar technique
has been used in \refcite{ham11:224305} for converting an
MCTDH wavefunction into ML-MCTDH format.
The technically-minded reader may find a detailed specification of
the MLPF algorithm in \refapp{app:mlpf}.
Here, it shall be sufficient to illustrate this procedure
using a system with 8 DOFs which are organized into a binary tree of modes.

For simplicity, it is assumed that each of the 8 DOFs employs $N$ grid points.
MLPF starts like regular Potfit, that is,
first the PES function is evaluated on the full grid, yielding
the potential energy tensor $V_{\alpha_1 \cdots \alpha_8}$
($\alpha_f = 1 \ldots N$). Then, as in Potfit, this order-8 tensor is
subjected to HOSVD, which yields an order-8 core tensor
$C_{i_1 \cdots i_8}$ plus, for each DOF, the $m$ dominant
natural potentials $v^\prim{f}_{i_f}$ ($i_f = 1 \cdots m$) --
again for simplicity, here it is assumed that $m$ is the same for all DOFs.
This gives a first approximation to $V$:
\begin{align}
V_{\alpha_1 \cdots \alpha_8} \approx \sum_{i_1=1}^{m} \cdots 
\sum_{i_8=1}^{m} C_{i_1 \cdots i_8} v^\prim{1}_{i_1 \alpha_1}
\cdots v^\prim{8}_{i_8 \alpha_8}
\label{eq:mlpf-core1}
\end{align}
Next, two DOFs each are combined into logical modes,
and the corresponding multi-indices
\begin{align}
\beta_1 &= (i_1,i_2)  &  \beta_2 &= (i_3,i_4)  &
\beta_3 &= (i_5,i_6)  &  \beta_4 &= (i_7,i_8)
\nonumber
\end{align}
are introduced. Then the order-8 core tensor is reinterpreted
or \emph{reshaped} into an order-4 tensor,
\begin{align}
C_{i_1 \cdots i_8} \equiv C_{\beta_1 \cdots \beta_4}
\quad(\beta_\kappa = 1 \ldots m^2)
\:.
\end{align}
Next, this reshaped core tensor is subjected to a second HOSVD,
which yields a second order-4 core tensor $\bar{C}_{j_1 \cdots j_4}$
plus, for each mode, the $\bar{m}$ dominant natural potentials
$\bar{v}^{(\kappa)}_{j_\kappa}$ ($j_\kappa = 1 \ldots \bar{m}$).
If $\bar{m} < m^2$, this results in an approximation to $C$:
\begin{align}
C_{\beta_1 \cdots \beta_4} \approx
\sum_{j_1=1}^{\bar{m}} \cdots \sum_{j_4=1}^{\bar{m}}
\bar{C}_{j_1 \cdots j_4} \bar{v}^{(1)}_{j_1 \beta_1} \cdots \bar{v}^{(4)}_{j_4 \beta_4}
\label{eq:mlpf-core2}
\end{align}
This process is now repeated. Two logical modes each are
combined into meta-modes, and the multi-indices
\begin{align}
\gamma_1 &= (j_1,j_2) & \gamma_2 &= (j_3,j_4) \nonumber
\end{align}
are introduced. The order-4 tensor $\bar{C}$ is reshaped into
an order-2 tensor,
\begin{align}
\bar{C}_{j_1 j_2 j_3 j_4} \equiv \bar{C}_{\gamma_1 \gamma_2}
\quad(\gamma_\mu = 1 \ldots \bar{m}^2)
\:,
\end{align}
and subjected to a third HOSVD (which now happens to be
a regular matrix SVD), yielding another order-2 core
tensor $\bbar{C}_{k_1 k_2}$ and natural potentials
$\bbar{v}^{(\mu)}_{k_\mu}$ ($k_\mu = 1 \ldots \bbar{m}$).
If $\bbar{m} < \bar{m}^2$, this results in an approximation
to $\bar{C}$:
\begin{align}
\bar{C}_{\gamma_1 \gamma_2} \approx
\sum_{k_1=1}^{\bbar{m}} \sum_{k_2=1}^{\bbar{m}}
\bbar{C}_{k_1 k_2} \bbar{v}^{(1)}_{k_1 \gamma_1} \bbar{v}^{(2)}_{k_2 \gamma_2}
\label{eq:mlpf-core3}
\end{align}
This completes the MLPF procedure. The resulting data are the
natural potentials ($v^\prim{f}$, $\bar{v}^{(\kappa)}$, $\bbar{v}^{(\mu)}$)
and the final core tensor $\bbar{C}$.

In order to connect the output of MLPF with the multi-layer operator structure
introduced in \refsec{sec:mlop}, one first associates (as in Potfit) the natural
potentials for the primitive modes with one-dimensional local operators $\hat{v}^\prim{f}_i$,
such that $\langle \chi^\prim{f}_\alpha | \hat{v}^\prim{f}_i | \chi^\prim{f}_{\alpha'} \rangle = v^\prim{f}_{i \alpha} \delta_{\alpha \alpha'}$.
Inserting Eqs.\,(\ref{eq:mlpf-core3}) and (\ref{eq:mlpf-core2}) into \refeq{eq:mlpf-core1}
and rearranging the summations, one then arrives at the following approximation
for the potential energy operator:\\

\begin{widetext}
\begin{align}
\hat{V} \approx
\sum_{k_1=1}^{\bbar{m}} \sum_{k_2=1}^{\bbar{m}} \bbar{C}_{k_1 k_2}
& \left[
    \sum_{j_1=1}^{\bar{m}} \sum_{j_2=1}^{\bar{m}} \bbar{v}^{(1)}_{k_1, j_1 j_2}
    \left(
      \sum_{i_1=1}^{m} \sum_{i_2=1}^{m} \bar{v}^{(1)}_{j_1,i_1 i_2}
      \hat{v}^\prim{1}_{i_1} \otimes \hat{v}^\prim{2}_{i_2}
    \right) \otimes
    \left(
      \sum_{i_3=1}^{m} \sum_{i_4=1}^{m} \bar{v}^{(2)}_{j_2,i_3 i_4}
      \hat{v}^\prim{3}_{i_3} \otimes \hat{v}^\prim{4}_{i_4} 
    \right)
  \right] \nonumber
\\ \otimes\: &
  \left[
    \sum_{j_3=1}^{\bar{m}} \sum_{j_4=1}^{\bar{m}} \bbar{v}^{(2)}_{k_2, j_3 j_4} 
    \left(
      \sum_{i_5=1}^{m} \sum_{i_6=1}^{m} \bar{v}^{(3)}_{j_3,i_5 i_6}
      \hat{v}^\prim{5}_{i_5} \otimes \hat{v}^\prim{6}_{i_6}
    \right) \otimes
    \left(
      \sum_{i_7=1}^{m} \sum_{i_8=1}^{m} \bar{v}^{(4)}_{j_4,i_7 i_8}
      \hat{v}^\prim{7}_{i_7} \otimes \hat{v}^\prim{8}_{i_8} 
    \right)
  \right]
\label{eq:mlpf3L}
\end{align}
\end{widetext}

In this form, the multi-layer operator structure is rather evident.
A direct comparison with \refeqq{def:hierop}{def:hierop2} reveals
the following associations:
\begin{align*}
V^\topn &= \bbar{C} \\
V^{(1)} &= \bbar{v}^{(1)} \qquad \mkern12mu V^{(2)} = \bbar{v}^{(2)} \\
V^{(1,1)} &= \bar{v}^{(1)} \qquad V^{(1,2)} = \bar{v}^{(2)} \\
V^{(2,1)} &= \bar{v}^{(3)} \qquad V^{(2,2)} = \bar{v}^{(4)} \\
\hat{U}^\prim{f}_b &= \hat{v}^\prim{f}_b \quad(f=1 \ldots 8,\; b=1 \ldots m)
\end{align*}
That is, the top-level transfer tensor is given by the final
core tensor, the other transfer tensors are given
by the natural potentials of the logical \mbox{(meta-)modes},
and the leaf operators are given by the natural potentials of the
primitive modes.
\refeq{eq:mlpf3L} also contains the expressions for the non-leaf single particle operators:
the four terms in parentheses give the SPOs $\hat{U}^{(1,1)}_{j_1}$, $\hat{U}^{(1,2)}_{j_2}$,
$\hat{U}^{(2,1)}_{j_3}$, and $\hat{U}^{(2,2)}_{j_4}$, respectively, whereas the two terms in
square brackets yield the SPOs $\hat{U}^{(1)}_{k_1}$ and $\hat{U}^{(2)}_{k_2}$.

The quality of the multi-layer approximation $\tilde{V}$ created by MLPF depends, of course,
on the number of natural potentials that are kept in each step, the
so-called \emph{truncation ranks} ($m, \bar{m}, \bbar{m}$ in the
above example).  Larger truncation ranks lead to a more accurate
approximation.  In fact, the error is strictly bounded by the sum of all neglected
natural weights (see Theorem 11.64 in \refcite{hac12}\footnote{%
Note that in the current edition, the squares are missing from the singular values
(W. Hackbusch, private communication). The squares of the singular values are
equal to the natural weights.}%
), i.e.
\begin{align}
\| V - \tilde{V} \|^2 \leq \sum_{z \neq \topn} \, \sum_{i > m_z} \lambda^{(z)}_i
\:.
\label{eq:hiersvd-error}
\end{align}
This is the multi-layer analogue of the error estimate for Potfit, \refeq{eq:pf-error}.
Additionally, it can be shown (ibid.) that the MLPF approximation is close to optimal, namely
\begin{align}
\| V - \tilde{V} \|^2 \leq (K-1) \| V - V_\text{best} \|^2
\label{eq:hiersvd-opt}
\end{align}
where $K$ is the number of nodes in the multi-layer tree ($K=2d-1$ for a binary tree),
and $V_\text{best}$ is the best possible MLPF approximation with the same
tree structure and the same truncation ranks, i.e. the best possible
approximation with the same format and size as $\tilde{V}$.
(The equivalent of \refeq{eq:hiersvd-opt} for regular Potfit is discussed in \refcite{pel13:014108}.)
\refeq{eq:hiersvd-opt} can be slightly improved for the case that
the top node has only two children; then the final
HOSVD is just a regular SVD, which has a better error bound,
and $(K-1)$ can be reduced to $(K-2)$.

\refeq{eq:hiersvd-error} opens a path for an \emph{error-controlled} truncation
strategy, i.e. one can determine the truncation ranks based on the desired accuracy
of the representation.  This is highly convenient, because the alternative manual
adjustment of the truncation ranks for many nodes would be
cumbersome and time-consuming.  \refapp{app:errctrl} presents one
possible implementation of such an error-controlled truncation scheme;
this scheme has been adopted for the computational example presented
in \refsec{sec:compex}.

Regarding the computational cost for the MLPF procedure, note that the core tensor becomes
smaller and smaller as the algorithm progresses. Hence the majority of the effort
is spent in the initial two steps (evaluating the PES on the full grid, and performing
the initial HOSVD). Consequently, the effort for MLPF is not much higher than
for Potfit itself. However, it is also not lower, and the same limiting factors that
affect Potfit also affect MLPF -- specifically, the need to evaluate the PES on the
full product grid. Options to mitigate this problem will be discussed in
\refsec{sec:mlpf-limitations}.

\section{Discussion}
\label{sec:discuss}

\subsection{Computational cost for ML-MCTDH}
\label{sec:numeff}

To see how much performance ML-MCTDH can gain by using
MLPF instead of Potfit for representing the potential, one must
estimate the numerical effort for both the sum-of-products
and the multi-layer operator formats.  
Such an estimate is simplified if a homogeneous structure
of the multi-layer tree is assumed.

Here we will consider a perfectly balanced tree with $L$ layers and mode
combinations of order $2$, such that there are $d=2^L$ primitive modes.
Furthermore, assume that the number of SPFs is $n_z=n$ for all leaf nodes,
and that it increases by a factor of $a$ when going from one layer to the
next-higher one (i.e. $n_z=an$ for the 1st layer above the bottom,
$n_z=a^2n$ for the 2nd, and so on---but $n_\topn=1$).
The same situation was considered by Manthe in \refcite{man08:164116}, where he
showed that the size of the wavefunction (i.e. the overall number of $A$-coefficients)
scales as $n^3 d^{3 \log_2 a}$ (for $a>2^{1/3}$), and it was suggested that the same
scaling applies to the computational cost for ML-MCTDH. A more detailed analysis is
presented in \refapp{app:numeff}. Here the results are summarized
and analyzed.

If, as in Potfit, a sum-of-products operator is used to represent the potential, the scaling of the ML-MCTDH
cost is given by \refeq{eq:totcost-sop}, and if each primitive mode employs $m$ natural
potentials so that Potfit needs $s=m^{d-1}$ operator terms (cf. \refeq{eq:contrpf}), then
the computational cost for evaluating the ML-MCTDH equations of motion for the whole tree scales as
\begin{align}
\text{cost(Potfit)} \propto m^{d-1} n^4 d^{4 \log_2 a}
\:.
\end{align}
For MLPF, a similar tree complexity as for the multi-layer wavefunction is assumed:
each primitive mode employs $m$ natural potentials, and this number increases by
a factor of $b$ when going one layer up. For this scenerio, 
the scaling of the ML-MCTDH cost is given by \refeq{eq:totcost-mlpf} as
\begin{align}
\text{cost(MLPF)} \propto m^2 n^4 d^{4 \log_2 a + 2 \log_2 b}
\quad\text{(if $a^2>b$)}
\:.
\end{align}
One sees that both potential representations lead to the
same dependence on the wavefunction complexity (i.e. on $n$ and $a$),
but the dependence on the operator complexity (i.e. on $m$ and (for MLPF) on $b$)
is vastly different: the cost using Potfit scales exponentially with $d$,
while the cost using MLPF scales only polynomially with $d$ (e.g. $\propto d^6$ for $a=b=2$).
The ratio of the computational costs scales as
\begin{align}
\frac{\text{cost(MLPF)}}{\text{cost(Potfit)}} \propto \frac{d^{2 \log_2 b}}{m^{d-3} }
\:,
\label{eq:costred}
\end{align}
i.e. MLPF is \emph{exponentially} cheaper than Potfit for increasing $d$.
It can also be seen that MLPF is especially beneficial for larger $m$, i.e. for
more accurate potential representations. The latter result already holds for $d=4$,
which is the minimum dimensionality for which the multi-layer approach
makes senses.

\begin{figure}
\includegraphics[scale=1.0]{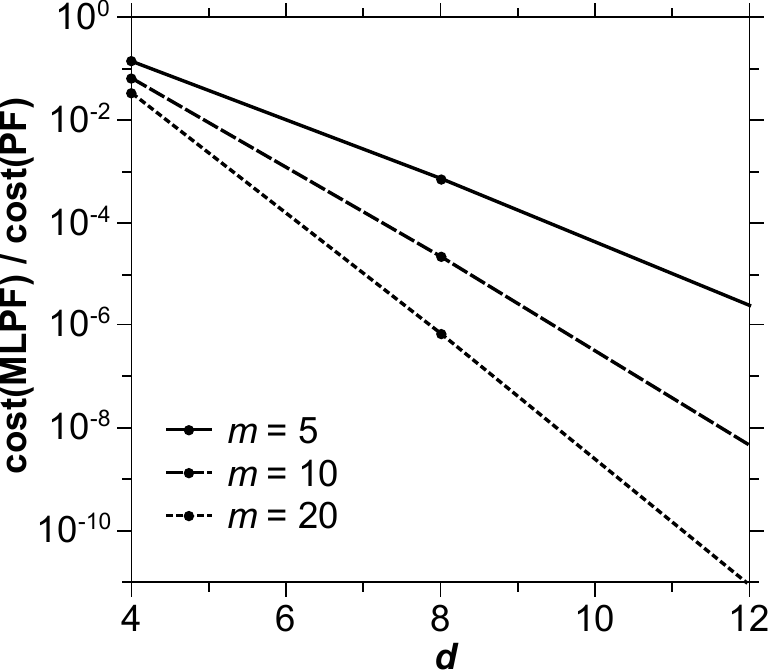}
\caption{Estimated savings in the computational cost for ML-MCTDH when using a
potential in MLPF instead of Potfit representation. $d$ is the number of 
primitive modes. The ML-tree is as described in the text, with
$n=10$ and $a=b=2$. Higher values of $m$ correspond to more accurate
potential representations. The values of $n$ and $a$ influence these
curves only marginally. Data was computed for $d=4$, $d=8$,
and $d=16$; the lines show interpolations to other values of $d$.}
\label{fig:savings}
\end{figure}

The asymptotic scaling behaviour discussed above shows that ML-MCTDH
can profit from using MLPF given that $d$ is ``large enough'', but for
practical purposes an actual estimate of this minimum $d$ would be
highly desirable. In fact, such an estimate can be rather easily obtained by
accurately adding up all the counts of arithmetic operations
for all nodes in the multi-layer tree (as detailed in \refapp{app:numeff}),
if specific values for $n$, $m$, $a$, and $b$ are chosen.
For $n=10$ and $a=b=2$, \reffig{fig:savings} shows how much the
ML-MCTDH cost can be reduced by using MLPF instead of Potfit,
for three different accuracy levels of the potential representation
($m=5$, $m=10$, $m=20$). It can be seen that already for $d=4$, there
is hope for reducing the computational cost by one order of magnitude
or more. As expected from the asymptotic scaling behaviour (cf. \refeq{eq:costred}),
the savings increase exponentially with $d$, and MLPF proves to be
especially advantageous for high-accuracy potential representations (i.e.
for large $m$).

\subsection{Limitations of multi-layer Potfit}
\label{sec:mlpf-limitations}

In \refsec{sec:mlop-eom}, it was assumed that the multi-layer operator
and the ML-MCTDH wavefunction have
identical tree structures. But it is quite common that a PES applies
only to a subset of the system's DOFs -- consider e.g. the situation of
two weakly interacting subsystems, where each subsystem has its own PES
and the interaction is described by coupling terms involving only
some of the DOFs. Using such a ``partial'' PES in MLPF will lead to
a multi-layer  operator that only acts on some of the DOFs, so obviously this
operator cannot have the same tree structure as the wavefunction. 
However, if the operator tree is identical to a \emph{subtree} of the
wavefunction, the operator can trivially be multiplied with unit operators
acting on the missing DOFs, and by using unit transfer tensors, it is
straightforward to build up an equivalent full multi-layer operator with
the correct tree structure.  In fact, this scheme also applies to any
product of a partial multi-layer operator with products of one-dimensional
operators, and even to direct products of multi-layer operators (as long as
they act on disjoint sets of DOFs). Hence in practice, the tree structure
used in MLPF must merely be \emph{compatible} with the tree structure
of the wavefunction (i.e. it should match some subtree), it doesn't actually
need to be identical.

As discussed in \refsec{sec:mlpf} and detailed in
\refapp{app:errctrl}, the accuracy of the MLPF representation
can be easily controlled by specifying the desired root-mean-square (RMS) error.
Unfortunately, this error measure is not always the most physically sensible.
First, the RMS error constitutes an average error, and as such it may not be
sensitive to very sharp spikes or dips. Missing or erroneously introducing such
features in the potential may however affect the wavefunction quite strongly.
Hence it would be desirable to at least also estimate the \emph{maximum}
error, but it's difficult to do so without running over the full grid.
Another problem is that the RMS error gives equal weight to all points of the
product grid, while one is often only interested in a ``relevant'' region of the
whole configuration space, e.g. those regions where the potential energy lies
below a certain threshold so that such regions are unreachable by the
wavefunction. This can be modeled by introducing (continuous or binary)
weights for the grid points, and then trying to minimize the \emph{weighted}
RMS error. The Potfit algorithm contains an option for such a minimization
through an iterative procedure, and as the first two steps of MLPF are basically
equivalent to Potfit, it should be easy to incorporate a similar feature into
MLPF. However, this is not implemented in the current version of MLPF.

Like Potfit, MLPF starts by evaluating the PES on the full product grid.
As already discussed at the end of \refsec{sec:potfit}, this limits the
applicability of the method, as the size of the product grid scales exponentially
with the number of primitive modes.  But as with Potfit, the multigrid Potfit (MGPF) method%
\cite{pel13:014108} may also be used here to alleviate this problem. In fact, it is
straightforward to take the potential representation returned by MGPF, compute
the natural potentials for the contracted mode, and then start the MLPF
algorithm with the core tensor (i.e. omitting the first two steps). This approach
has already been tried successfully on a system with 9 DOFs, which will be the subject of a separate publication.
Combining MLPF with MGPF can probably extend its applicability to PES's
with up to 12 DOFs. For larger systems, it will likely be necessary to develop
new methods for fitting the PES into MLPF-format.

\section{A computational example}
\label{sec:compex}

In this section, the benefits of MLPF shall be demonstrated using
a realistic example. The example system to be considered is the inelastic
collision of two H$_2$ molecules. As our group has studied this system
in the past\cite{gat05:174311,pan07:114310,ott08:064305,ott09:049901,ott12:619}
using standard MCTDH, here a brief summary of the system details
is sufficient.

\begin{figure}
{\fontsize{9}{11}\selectfont
\begin{tikzpicture}[>=latex,line join=bevel, scale=0.55, every node/.style={scale=1}]
  \pgfsetlinewidth{0.6bp}
\pgfsetcolor{black}
  \draw [] (143.86bp,22bp) -- (119.57bp,67.894bp);
  \definecolor{strokecol}{rgb}{0.0,0.0,0.0};
  \pgfsetstrokecolor{strokecol}
  \draw (144.5bp,45bp) node {$11$};
  \draw [] (16.632bp,80.091bp) -- (62.603bp,127.33bp);
  \draw (57.5bp,103bp) node {$14$};
  \draw [] (173.22bp,79.328bp) -- (119.21bp,127.13bp);
  \draw (168.5bp,103bp) node {$20$};
  \draw [] (49.375bp,22bp) -- (63.276bp,66.608bp);
  \draw (63bp,43bp) node {$8$};
  \draw [] (10.17bp,22bp) -- (10.852bp,63.848bp);
  \draw (25bp,43bp) node {$192$};
  \draw [] (180bp,22bp) -- (180bp,63.632bp);
  \draw (189.5bp,41bp) node {$10$};
  \draw [] (112bp,82.113bp) -- (112bp,123.8bp);
  \draw (123.5bp,103bp) node {$20$};
  \draw [] (78.349bp,22bp) -- (71.49bp,63.76bp);
  \draw (85bp,43bp) node {$8$};
  \draw [] (213.97bp,22bp) -- (187.81bp,68.35bp);
  \draw (214.5bp,45bp) node {$11$};
  \draw [] (108.85bp,141.6bp) -- (93.146bp,184.42bp);
  \draw (115.5bp,163bp) node {$20$};
  \draw [] (70bp,82.113bp) -- (70bp,123.8bp);
  \draw (81.5bp,103bp) node {$20$};
  \draw [] (110.36bp,22bp) -- (111.7bp,63.632bp);
  \draw (120.5bp,41bp) node {$10$};
  \draw [] (70.555bp,142.18bp) -- (84.167bp,185.61bp);
  \draw (88.5bp,163bp) node {$20$};
\begin{scope}
  \definecolor{strokecol}{rgb}{0.0,0.0,0.0};
  \pgfsetstrokecolor{strokecol}
  \draw (20bp,22bp) -- (0bp,22bp) -- (0bp,1bp) -- (20bp,1bp) -- cycle;
  \draw (10bp,11bp) node {$R$};
\end{scope}
\begin{scope}
  \definecolor{strokecol}{rgb}{0.0,0.0,0.0};
  \pgfsetstrokecolor{strokecol}
  \draw (61bp,22bp) -- (39bp,22bp) -- (39bp,1bp) -- (61bp,1bp) -- cycle;
  \draw (50bp,11bp) node {$r_1$};
\end{scope}
\begin{scope}
  \definecolor{strokecol}{rgb}{0.0,0.0,0.0};
  \pgfsetstrokecolor{strokecol}
  \draw (91bp,22bp) -- (69bp,22bp) -- (69bp,1bp) -- (91bp,1bp) -- cycle;
  \draw (80bp,11bp) node {$r_2$};
\end{scope}
\begin{scope}
  \definecolor{strokecol}{rgb}{0.0,0.0,0.0};
  \pgfsetstrokecolor{strokecol}
  \draw (121bp,22bp) -- (99bp,22bp) -- (99bp,1bp) -- (121bp,1bp) -- cycle;
  \draw (110bp,11bp) node {$\theta_1$};
\end{scope}
\begin{scope}
  \definecolor{strokecol}{rgb}{0.0,0.0,0.0};
  \pgfsetstrokecolor{strokecol}
  \draw (156bp,22bp) -- (134bp,22bp) -- (134bp,1bp) -- (156bp,1bp) -- cycle;
  \draw (145bp,11bp) node {$k_1$};
\end{scope}
\begin{scope}
  \definecolor{strokecol}{rgb}{0.0,0.0,0.0};
  \pgfsetstrokecolor{strokecol}
  \draw (191bp,22bp) -- (169bp,22bp) -- (169bp,1bp) -- (191bp,1bp) -- cycle;
  \draw (180bp,11bp) node {$\theta_2$};
\end{scope}
\begin{scope}
  \definecolor{strokecol}{rgb}{0.0,0.0,0.0};
  \pgfsetstrokecolor{strokecol}
  \draw (226bp,22bp) -- (204bp,22bp) -- (204bp,1bp) -- (226bp,1bp) -- cycle;
  \draw (215bp,11bp) node {$k_2$};
\end{scope}
\begin{scope}
  \definecolor{strokecol}{rgb}{0.0,0.0,0.0};
  \pgfsetstrokecolor{strokecol}
  \draw (90bp,193bp) ellipse (9bp and 9bp);
  \draw (90bp,193bp) node {$ $};
\end{scope}
\begin{scope}
  \definecolor{strokecol}{rgb}{0.0,0.0,0.0};
  \pgfsetstrokecolor{strokecol}
  \draw (70bp,133bp) ellipse (9bp and 9bp);
  \draw (70bp,133bp) node {$ $};
\end{scope}
\begin{scope}
  \definecolor{strokecol}{rgb}{0.0,0.0,0.0};
  \pgfsetstrokecolor{strokecol}
  \draw (11bp,73bp) ellipse (9bp and 9bp);
  \draw (11bp,73bp) node {$ $};
\end{scope}
\begin{scope}
  \definecolor{strokecol}{rgb}{0.0,0.0,0.0};
  \pgfsetstrokecolor{strokecol}
  \draw (70bp,73bp) ellipse (9bp and 9bp);
  \draw (70bp,73bp) node {$ $};
\end{scope}
\begin{scope}
  \definecolor{strokecol}{rgb}{0.0,0.0,0.0};
  \pgfsetstrokecolor{strokecol}
  \draw (112bp,133bp) ellipse (9bp and 9bp);
  \draw (112bp,133bp) node {$ $};
\end{scope}
\begin{scope}
  \definecolor{strokecol}{rgb}{0.0,0.0,0.0};
  \pgfsetstrokecolor{strokecol}
  \draw (112bp,73bp) ellipse (9bp and 9bp);
  \draw (112bp,73bp) node {$ $};
\end{scope}
\begin{scope}
  \definecolor{strokecol}{rgb}{0.0,0.0,0.0};
  \pgfsetstrokecolor{strokecol}
  \draw (180bp,73bp) ellipse (9bp and 9bp);
  \draw (180bp,73bp) node {$ $};
\end{scope}
\end{tikzpicture}}
\caption{ML-MCTDH tree for the H$_2$+H$_2$ scattering
system described in the text.  The edges are labeled according
to the number of SPFs (upper two layers) or primitive basis
size (bottom layer).}
\label{fig:mlmctdh-tree}
\end{figure}
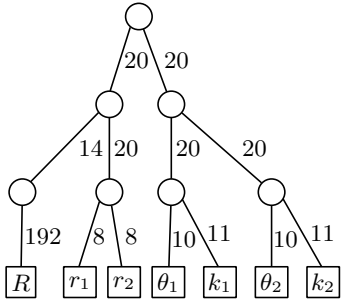

The system is described by six internal coordinates plus one additional
angle that results from using the so-called $\text{E}_2$ frame instead
of the body-fixed frame, which is convenient for total angular
momentum $J>0$.  Hence seven physical degrees of freedom are used:
the intermolecular distance $R$, the H$_2$ bond lengths $r_i$ ($i=1,2$),
and polar/azimuthal angles $\theta_i/\phi_i$ describing the orientation
of the molecules with respect to the intermolecular axis.
Due to the structure of the kinetic energy operator,
it is beneficial for the wavepacket propagation to use instead of the
angles $\phi_i$ their conjugate momenta $k_i$. This amounts to a
Fourier transform of the wavefunction in $\phi_1$ and $\phi_2$.
In the previous MCTDH calculations, four primitive modes were
employed: $R$, $(r_1,r_2)$, $(\theta_1,k_1)$, $(\theta_2,k_2)$. Now
for the ML-MCTDH calculations, the same primitive modes are used, and one
intermediate layer is introduced. The resulting tree structure is depicted
in \reffig{fig:mlmctdh-tree}, where also the numbers of primitive basis
functions (bottom layer) and the numbers of SPFs (upper two layers)
are shown. These SPF numbers were chosen to be rather high, i.e.
a rather accurate wavefunction representation was used, so that
inaccuracies in the potential representation are not obscured by
inaccuracies in the wavefunction.

In the present work, the potential energy surface of Boothroyd et al.%
\cite{boo02:666} for the $(\text{H}_2)_2$ system was used.  This ``BMKP'' PES
is given as a six-dimensional function, $V(R,r_1,r_2,\theta_1,\theta_2,\phi)$
where $\phi=\phi_1-\phi_2$. To use this PES in the wavepacket propagation,
one must perform a Fourier transform along $\phi$, leading to
five-dimensional Fourier components $V_\Omega(R,r_1,r_2,\theta_1,\theta_2)$
($\Omega=0,1,2,\ldots$). Keeping only components with $\Omega \leq 3$
proved sufficient for convergence (and only $\Omega \geq 0$ need to be
considered as here $V_\Omega = V_{-\Omega}$). Each Fourier component, initially
given on the primitive product grid of $1.23\cdot10^6$ points,
was then brought into Potfit- or MLPF-format. As this step \emph{truncates}
the potential representation, it constitutes an approximation.

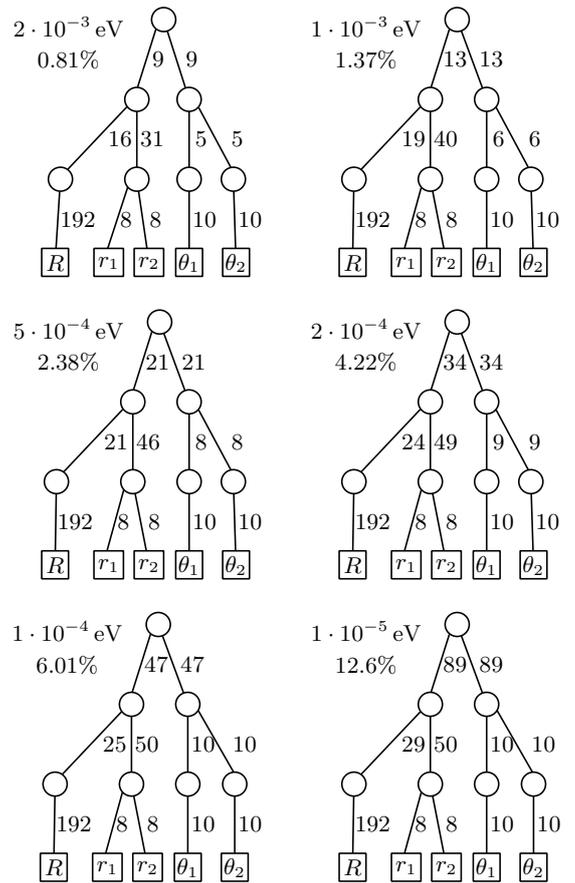
\begin{figure}
\fontsize{9}{11}\selectfont

\subfigure{
\begin{tikzpicture}[>=latex,line join=bevel, scale=0.5, every node/.style={scale=0.96}]
  \pgfsetlinewidth{0.6bp}
\pgfsetcolor{black}

\draw(20bp,190bp) node {$2 \cdot 10^{-3}\,\text{eV}$};
\draw(20bp,165bp) node {0.81\%};

  \draw [] (19.46bp,82.427bp) -- (63.553bp,129.38bp);
  \definecolor{strokecol}{rgb}{0.0,0.0,0.0};
  \pgfsetstrokecolor{strokecol}
  \draw (58.5bp,105bp) node {$16$};
  \draw [] (144.63bp,23bp) -- (143.29bp,65.965bp);
  \draw (155.5bp,45bp) node {$10$};
  \draw [] (49.254bp,23bp) -- (64.198bp,69.032bp);
  \draw (63bp,45bp) node {$8$};
  \draw [] (10.679bp,23bp) -- (13.41bp,65.7bp);
  \draw (27bp,45bp) node {$192$};
  \draw [] (110bp,84.113bp) -- (110bp,125.8bp);
  \draw (119bp,105bp) node {$5$};
  \draw [] (141bp,84.058bp) -- (117.1bp,128.92bp);
  \draw (146bp,105bp) node {$5$};
  \draw [] (78.554bp,23bp) -- (72.296bp,65.925bp);
  \draw (85bp,45bp) node {$8$};
  \draw [] (107.28bp,143.6bp) -- (93.717bp,186.42bp);
  \draw (112bp,165bp) node {$9$};
  \draw [] (71bp,84.113bp) -- (71bp,125.8bp);
  \draw (82.5bp,105bp) node {$31$};
  \draw [] (110bp,23bp) -- (110bp,65.99bp);
  \draw (121.5bp,45bp) node {$10$};
  \draw [] (71.555bp,144.18bp) -- (85.167bp,187.61bp);
  \draw (87bp,165bp) node {$9$};
\begin{scope}
  \definecolor{strokecol}{rgb}{0.0,0.0,0.0};
  \pgfsetstrokecolor{strokecol}
  \draw (20bp,23bp) -- (0bp,23bp) -- (0bp,2bp) -- (20bp,2bp) -- cycle;
  \draw (10bp,12bp) node {$R$};
\end{scope}
\begin{scope}
  \definecolor{strokecol}{rgb}{0.0,0.0,0.0};
  \pgfsetstrokecolor{strokecol}
  \draw (61bp,23bp) -- (39bp,23bp) -- (39bp,2bp) -- (61bp,2bp) -- cycle;
  \draw (50bp,12bp) node {$r_1$};
\end{scope}
\begin{scope}
  \definecolor{strokecol}{rgb}{0.0,0.0,0.0};
  \pgfsetstrokecolor{strokecol}
  \draw (91bp,23bp) -- (69bp,23bp) -- (69bp,2bp) -- (91bp,2bp) -- cycle;
  \draw (80bp,12bp) node {$r_2$};
\end{scope}
\begin{scope}
  \definecolor{strokecol}{rgb}{0.0,0.0,0.0};
  \pgfsetstrokecolor{strokecol}
  \draw (120bp,23bp) -- (100bp,23bp) -- (100bp,2bp) -- (120bp,2bp) -- cycle;
  \draw (110bp,12bp) node {$\theta_1$};
\end{scope}
\begin{scope}
  \definecolor{strokecol}{rgb}{0.0,0.0,0.0};
  \pgfsetstrokecolor{strokecol}
  \draw (155bp,23bp) -- (135bp,23bp) -- (135bp,2bp) -- (155bp,2bp) -- cycle;
  \draw (145bp,12bp) node {$\theta_2$};
\end{scope}
\begin{scope}
  \definecolor{strokecol}{rgb}{0.0,0.0,0.0};
  \pgfsetstrokecolor{strokecol}
  \draw (91bp,195bp) ellipse (9bp and 9bp);
  \draw (91bp,195bp) node {$$};
\end{scope}
\begin{scope}
  \definecolor{strokecol}{rgb}{0.0,0.0,0.0};
  \pgfsetstrokecolor{strokecol}
  \draw (71bp,135bp) ellipse (9bp and 9bp);
  \draw (71bp,135bp) node {$$};
\end{scope}
\begin{scope}
  \definecolor{strokecol}{rgb}{0.0,0.0,0.0};
  \pgfsetstrokecolor{strokecol}
  \draw (14bp,75bp) ellipse (9bp and 9bp);
  \draw (14bp,75bp) node {$$};
\end{scope}
\begin{scope}
  \definecolor{strokecol}{rgb}{0.0,0.0,0.0};
  \pgfsetstrokecolor{strokecol}
  \draw (71bp,75bp) ellipse (9bp and 9bp);
  \draw (71bp,75bp) node {$$};
\end{scope}
\begin{scope}
  \definecolor{strokecol}{rgb}{0.0,0.0,0.0};
  \pgfsetstrokecolor{strokecol}
  \draw (110bp,135bp) ellipse (9bp and 9bp);
  \draw (110bp,135bp) node {$$};
\end{scope}
\begin{scope}
  \definecolor{strokecol}{rgb}{0.0,0.0,0.0};
  \pgfsetstrokecolor{strokecol}
  \draw (110bp,75bp) ellipse (9bp and 9bp);
  \draw (110bp,75bp) node {$$};
\end{scope}
\begin{scope}
  \definecolor{strokecol}{rgb}{0.0,0.0,0.0};
  \pgfsetstrokecolor{strokecol}
  \draw (143bp,75bp) ellipse (9bp and 9bp);
  \draw (143bp,75bp) node {$$};
\end{scope}
\end{tikzpicture}
}
\subfigure{
\begin{tikzpicture}[>=latex,line join=bevel, scale=0.5, every node/.style={scale=0.96}]
  \pgfsetlinewidth{0.6bp}
\pgfsetcolor{black}

\draw(20bp,190bp) node {$1 \cdot 10^{-3}\,\text{eV}$};
\draw(20bp,165bp) node {1.37\%};

  \draw [] (16.46bp,82.427bp) -- (60.553bp,129.38bp);
  \definecolor{strokecol}{rgb}{0.0,0.0,0.0};
  \pgfsetstrokecolor{strokecol}
  \draw (55.5bp,105bp) node {$19$};
  \draw [] (144.63bp,23bp) -- (143.29bp,65.965bp);
  \draw (155.5bp,45bp) node {$10$};
  \draw [] (48.738bp,23bp) -- (61.177bp,68.646bp);
  \draw (61bp,45bp) node {$8$};
  \draw [] (10.17bp,23bp) -- (10.852bp,65.7bp);
  \draw (25bp,45bp) node {$192$};
  \draw [] (110bp,84.113bp) -- (110bp,125.8bp);
  \draw (119bp,105bp) node {$6$};
  \draw [] (141bp,84.058bp) -- (117.1bp,128.92bp);
  \draw (146bp,105bp) node {$6$};
  \draw [] (78.072bp,23bp) -- (69.728bp,65.925bp);
  \draw (84bp,45bp) node {$8$};
  \draw [] (106.85bp,143.6bp) -- (91.146bp,186.42bp);
  \draw (113.5bp,165bp) node {$13$};
  \draw [] (68bp,84.113bp) -- (68bp,125.8bp);
  \draw (79.5bp,105bp) node {$40$};
  \draw [] (110bp,23bp) -- (110bp,65.99bp);
  \draw (121.5bp,45bp) node {$10$};
  \draw [] (68.555bp,144.18bp) -- (82.167bp,187.61bp);
  \draw (86.5bp,165bp) node {$13$};
\begin{scope}
  \definecolor{strokecol}{rgb}{0.0,0.0,0.0};
  \pgfsetstrokecolor{strokecol}
  \draw (20bp,23bp) -- (0bp,23bp) -- (0bp,2bp) -- (20bp,2bp) -- cycle;
  \draw (10bp,12bp) node {$R$};
\end{scope}
\begin{scope}
  \definecolor{strokecol}{rgb}{0.0,0.0,0.0};
  \pgfsetstrokecolor{strokecol}
  \draw (61bp,23bp) -- (39bp,23bp) -- (39bp,2bp) -- (61bp,2bp) -- cycle;
  \draw (50bp,12bp) node {$r_1$};
\end{scope}
\begin{scope}
  \definecolor{strokecol}{rgb}{0.0,0.0,0.0};
  \pgfsetstrokecolor{strokecol}
  \draw (91bp,23bp) -- (69bp,23bp) -- (69bp,2bp) -- (91bp,2bp) -- cycle;
  \draw (80bp,12bp) node {$r_2$};
\end{scope}
\begin{scope}
  \definecolor{strokecol}{rgb}{0.0,0.0,0.0};
  \pgfsetstrokecolor{strokecol}
  \draw (120bp,23bp) -- (100bp,23bp) -- (100bp,2bp) -- (120bp,2bp) -- cycle;
  \draw (110bp,12bp) node {$\theta_1$};
\end{scope}
\begin{scope}
  \definecolor{strokecol}{rgb}{0.0,0.0,0.0};
  \pgfsetstrokecolor{strokecol}
  \draw (155bp,23bp) -- (135bp,23bp) -- (135bp,2bp) -- (155bp,2bp) -- cycle;
  \draw (145bp,12bp) node {$\theta_2$};
\end{scope}
\begin{scope}
  \definecolor{strokecol}{rgb}{0.0,0.0,0.0};
  \pgfsetstrokecolor{strokecol}
  \draw (88bp,195bp) ellipse (9bp and 9bp);
  \draw (88bp,195bp) node {$ $};
\end{scope}
\begin{scope}
  \definecolor{strokecol}{rgb}{0.0,0.0,0.0};
  \pgfsetstrokecolor{strokecol}
  \draw (68bp,135bp) ellipse (9bp and 9bp);
  \draw (68bp,135bp) node {$ $};
\end{scope}
\begin{scope}
  \definecolor{strokecol}{rgb}{0.0,0.0,0.0};
  \pgfsetstrokecolor{strokecol}
  \draw (11bp,75bp) ellipse (9bp and 9bp);
  \draw (11bp,75bp) node {$ $};
\end{scope}
\begin{scope}
  \definecolor{strokecol}{rgb}{0.0,0.0,0.0};
  \pgfsetstrokecolor{strokecol}
  \draw (68bp,75bp) ellipse (9bp and 9bp);
  \draw (68bp,75bp) node {$ $};
\end{scope}
\begin{scope}
  \definecolor{strokecol}{rgb}{0.0,0.0,0.0};
  \pgfsetstrokecolor{strokecol}
  \draw (110bp,135bp) ellipse (9bp and 9bp);
  \draw (110bp,135bp) node {$ $};
\end{scope}
\begin{scope}
  \definecolor{strokecol}{rgb}{0.0,0.0,0.0};
  \pgfsetstrokecolor{strokecol}
  \draw (110bp,75bp) ellipse (9bp and 9bp);
  \draw (110bp,75bp) node {$ $};
\end{scope}
\begin{scope}
  \definecolor{strokecol}{rgb}{0.0,0.0,0.0};
  \pgfsetstrokecolor{strokecol}
  \draw (143bp,75bp) ellipse (9bp and 9bp);
  \draw (143bp,75bp) node {$ $};
\end{scope}
\end{tikzpicture}
}

\subfigure{
\begin{tikzpicture}[>=latex,line join=bevel, scale=0.5, every node/.style={scale=0.96}]
  \pgfsetlinewidth{0.6bp}
\pgfsetcolor{black}

\draw(20bp,190bp) node {$5 \cdot 10^{-4}\,\text{eV}$};
\draw(20bp,165bp) node {2.38\%};

  \draw [] (16.46bp,82.427bp) -- (60.553bp,129.38bp);
  \definecolor{strokecol}{rgb}{0.0,0.0,0.0};
  \pgfsetstrokecolor{strokecol}
  \draw (55.5bp,105bp) node {$21$};
  \draw [] (144.63bp,23bp) -- (143.29bp,65.965bp);
  \draw (155.5bp,45bp) node {$10$};
  \draw [] (48.738bp,23bp) -- (61.177bp,68.646bp);
  \draw (61bp,45bp) node {$8$};
  \draw [] (10.17bp,23bp) -- (10.852bp,65.7bp);
  \draw (25bp,45bp) node {$192$};
  \draw [] (110bp,84.113bp) -- (110bp,125.8bp);
  \draw (119bp,105bp) node {$8$};
  \draw [] (141bp,84.058bp) -- (117.1bp,128.92bp);
  \draw (146bp,105bp) node {$8$};
  \draw [] (78.072bp,23bp) -- (69.728bp,65.925bp);
  \draw (84bp,45bp) node {$8$};
  \draw [] (106.85bp,143.6bp) -- (91.146bp,186.42bp);
  \draw (113.5bp,165bp) node {$21$};
  \draw [] (68bp,84.113bp) -- (68bp,125.8bp);
  \draw (79.5bp,105bp) node {$46$};
  \draw [] (110bp,23bp) -- (110bp,65.99bp);
  \draw (121.5bp,45bp) node {$10$};
  \draw [] (68.555bp,144.18bp) -- (82.167bp,187.61bp);
  \draw (86.5bp,165bp) node {$21$};
\begin{scope}
  \definecolor{strokecol}{rgb}{0.0,0.0,0.0};
  \pgfsetstrokecolor{strokecol}
  \draw (20bp,23bp) -- (0bp,23bp) -- (0bp,2bp) -- (20bp,2bp) -- cycle;
  \draw (10bp,12bp) node {$R$};
\end{scope}
\begin{scope}
  \definecolor{strokecol}{rgb}{0.0,0.0,0.0};
  \pgfsetstrokecolor{strokecol}
  \draw (61bp,23bp) -- (39bp,23bp) -- (39bp,2bp) -- (61bp,2bp) -- cycle;
  \draw (50bp,12bp) node {$r_1$};
\end{scope}
\begin{scope}
  \definecolor{strokecol}{rgb}{0.0,0.0,0.0};
  \pgfsetstrokecolor{strokecol}
  \draw (91bp,23bp) -- (69bp,23bp) -- (69bp,2bp) -- (91bp,2bp) -- cycle;
  \draw (80bp,12bp) node {$r_2$};
\end{scope}
\begin{scope}
  \definecolor{strokecol}{rgb}{0.0,0.0,0.0};
  \pgfsetstrokecolor{strokecol}
  \draw (120bp,23bp) -- (100bp,23bp) -- (100bp,2bp) -- (120bp,2bp) -- cycle;
  \draw (110bp,12bp) node {$\theta_1$};
\end{scope}
\begin{scope}
  \definecolor{strokecol}{rgb}{0.0,0.0,0.0};
  \pgfsetstrokecolor{strokecol}
  \draw (155bp,23bp) -- (135bp,23bp) -- (135bp,2bp) -- (155bp,2bp) -- cycle;
  \draw (145bp,12bp) node {$\theta_2$};
\end{scope}
\begin{scope}
  \definecolor{strokecol}{rgb}{0.0,0.0,0.0};
  \pgfsetstrokecolor{strokecol}
  \draw (88bp,195bp) ellipse (9bp and 9bp);
  \draw (88bp,195bp) node {$$};
\end{scope}
\begin{scope}
  \definecolor{strokecol}{rgb}{0.0,0.0,0.0};
  \pgfsetstrokecolor{strokecol}
  \draw (68bp,135bp) ellipse (9bp and 9bp);
  \draw (68bp,135bp) node {$$};
\end{scope}
\begin{scope}
  \definecolor{strokecol}{rgb}{0.0,0.0,0.0};
  \pgfsetstrokecolor{strokecol}
  \draw (11bp,75bp) ellipse (9bp and 9bp);
  \draw (11bp,75bp) node {$$};
\end{scope}
\begin{scope}
  \definecolor{strokecol}{rgb}{0.0,0.0,0.0};
  \pgfsetstrokecolor{strokecol}
  \draw (68bp,75bp) ellipse (9bp and 9bp);
  \draw (68bp,75bp) node {$$};
\end{scope}
\begin{scope}
  \definecolor{strokecol}{rgb}{0.0,0.0,0.0};
  \pgfsetstrokecolor{strokecol}
  \draw (110bp,135bp) ellipse (9bp and 9bp);
  \draw (110bp,135bp) node {$$};
\end{scope}
\begin{scope}
  \definecolor{strokecol}{rgb}{0.0,0.0,0.0};
  \pgfsetstrokecolor{strokecol}
  \draw (110bp,75bp) ellipse (9bp and 9bp);
  \draw (110bp,75bp) node {$$};
\end{scope}
\begin{scope}
  \definecolor{strokecol}{rgb}{0.0,0.0,0.0};
  \pgfsetstrokecolor{strokecol}
  \draw (143bp,75bp) ellipse (9bp and 9bp);
  \draw (143bp,75bp) node {$$};
\end{scope}
\end{tikzpicture}
}
\subfigure{
\begin{tikzpicture}[>=latex,line join=bevel, scale=0.5, every node/.style={scale=0.96}]
  \pgfsetlinewidth{0.6bp}
\pgfsetcolor{black}

\draw(20bp,190bp) node {$2 \cdot 10^{-4}\,\text{eV}$};
\draw(20bp,165bp) node {4.22\%};

  \draw [] (16.46bp,82.427bp) -- (60.553bp,129.38bp);
  \definecolor{strokecol}{rgb}{0.0,0.0,0.0};
  \pgfsetstrokecolor{strokecol}
  \draw (55.5bp,105bp) node {$24$};
  \draw [] (144.63bp,23bp) -- (143.29bp,65.965bp);
  \draw (155.5bp,45bp) node {$10$};
  \draw [] (48.738bp,23bp) -- (61.177bp,68.646bp);
  \draw (61bp,45bp) node {$8$};
  \draw [] (10.17bp,23bp) -- (10.852bp,65.7bp);
  \draw (25bp,45bp) node {$192$};
  \draw [] (110bp,84.113bp) -- (110bp,125.8bp);
  \draw (119bp,105bp) node {$9$};
  \draw [] (141bp,84.058bp) -- (117.1bp,128.92bp);
  \draw (146bp,105bp) node {$9$};
  \draw [] (78.072bp,23bp) -- (69.728bp,65.925bp);
  \draw (84bp,45bp) node {$8$};
  \draw [] (106.85bp,143.6bp) -- (91.146bp,186.42bp);
  \draw (113.5bp,165bp) node {$34$};
  \draw [] (68bp,84.113bp) -- (68bp,125.8bp);
  \draw (79.5bp,105bp) node {$49$};
  \draw [] (110bp,23bp) -- (110bp,65.99bp);
  \draw (121.5bp,45bp) node {$10$};
  \draw [] (68.555bp,144.18bp) -- (82.167bp,187.61bp);
  \draw (86.5bp,165bp) node {$34$};
\begin{scope}
  \definecolor{strokecol}{rgb}{0.0,0.0,0.0};
  \pgfsetstrokecolor{strokecol}
  \draw (20bp,23bp) -- (0bp,23bp) -- (0bp,2bp) -- (20bp,2bp) -- cycle;
  \draw (10bp,12bp) node {$R$};
\end{scope}
\begin{scope}
  \definecolor{strokecol}{rgb}{0.0,0.0,0.0};
  \pgfsetstrokecolor{strokecol}
  \draw (61bp,23bp) -- (39bp,23bp) -- (39bp,2bp) -- (61bp,2bp) -- cycle;
  \draw (50bp,12bp) node {$r_1$};
\end{scope}
\begin{scope}
  \definecolor{strokecol}{rgb}{0.0,0.0,0.0};
  \pgfsetstrokecolor{strokecol}
  \draw (91bp,23bp) -- (69bp,23bp) -- (69bp,2bp) -- (91bp,2bp) -- cycle;
  \draw (80bp,12bp) node {$r_2$};
\end{scope}
\begin{scope}
  \definecolor{strokecol}{rgb}{0.0,0.0,0.0};
  \pgfsetstrokecolor{strokecol}
  \draw (120bp,23bp) -- (100bp,23bp) -- (100bp,2bp) -- (120bp,2bp) -- cycle;
  \draw (110bp,12bp) node {$\theta_1$};
\end{scope}
\begin{scope}
  \definecolor{strokecol}{rgb}{0.0,0.0,0.0};
  \pgfsetstrokecolor{strokecol}
  \draw (155bp,23bp) -- (135bp,23bp) -- (135bp,2bp) -- (155bp,2bp) -- cycle;
  \draw (145bp,12bp) node {$\theta_2$};
\end{scope}
\begin{scope}
  \definecolor{strokecol}{rgb}{0.0,0.0,0.0};
  \pgfsetstrokecolor{strokecol}
  \draw (88bp,195bp) ellipse (9bp and 9bp);
  \draw (88bp,195bp) node {$$};
\end{scope}
\begin{scope}
  \definecolor{strokecol}{rgb}{0.0,0.0,0.0};
  \pgfsetstrokecolor{strokecol}
  \draw (68bp,135bp) ellipse (9bp and 9bp);
  \draw (68bp,135bp) node {$$};
\end{scope}
\begin{scope}
  \definecolor{strokecol}{rgb}{0.0,0.0,0.0};
  \pgfsetstrokecolor{strokecol}
  \draw (11bp,75bp) ellipse (9bp and 9bp);
  \draw (11bp,75bp) node {$$};
\end{scope}
\begin{scope}
  \definecolor{strokecol}{rgb}{0.0,0.0,0.0};
  \pgfsetstrokecolor{strokecol}
  \draw (68bp,75bp) ellipse (9bp and 9bp);
  \draw (68bp,75bp) node {$$};
\end{scope}
\begin{scope}
  \definecolor{strokecol}{rgb}{0.0,0.0,0.0};
  \pgfsetstrokecolor{strokecol}
  \draw (110bp,135bp) ellipse (9bp and 9bp);
  \draw (110bp,135bp) node {$$};
\end{scope}
\begin{scope}
  \definecolor{strokecol}{rgb}{0.0,0.0,0.0};
  \pgfsetstrokecolor{strokecol}
  \draw (110bp,75bp) ellipse (9bp and 9bp);
  \draw (110bp,75bp) node {$$};
\end{scope}
\begin{scope}
  \definecolor{strokecol}{rgb}{0.0,0.0,0.0};
  \pgfsetstrokecolor{strokecol}
  \draw (143bp,75bp) ellipse (9bp and 9bp);
  \draw (143bp,75bp) node {$$};
\end{scope}
\end{tikzpicture}
}

\subfigure{
\begin{tikzpicture}[>=latex,line join=bevel, scale=0.5, every node/.style={scale=0.96}]
  \pgfsetlinewidth{0.6bp}
\pgfsetcolor{black}

\draw(20bp,190bp) node {$1 \cdot 10^{-4}\,\text{eV}$};
\draw(20bp,165bp) node {6.01\%};

  \draw [] (16.46bp,82.427bp) -- (60.553bp,129.38bp);
  \definecolor{strokecol}{rgb}{0.0,0.0,0.0};
  \pgfsetstrokecolor{strokecol}
  \draw (55.5bp,105bp) node {$25$};
  \draw [] (145bp,23bp) -- (145bp,65.965bp);
  \draw (156.5bp,45bp) node {$10$};
  \draw [] (48.738bp,23bp) -- (61.177bp,68.646bp);
  \draw (61bp,45bp) node {$8$};
  \draw [] (10.17bp,23bp) -- (10.852bp,65.7bp);
  \draw (25bp,45bp) node {$192$};
  \draw [] (110bp,84.113bp) -- (110bp,125.8bp);
  \draw (121.5bp,105bp) node {$10$};
  \draw [] (144.06bp,84.218bp) -- (117.78bp,130.33bp);
  \draw (152.5bp,105bp) node {$10$};
  \draw [] (78.072bp,23bp) -- (69.728bp,65.925bp);
  \draw (84bp,45bp) node {$8$};
  \draw [] (106.85bp,143.6bp) -- (91.146bp,186.42bp);
  \draw (113.5bp,165bp) node {$47$};
  \draw [] (68bp,84.113bp) -- (68bp,125.8bp);
  \draw (79.5bp,105bp) node {$50$};
  \draw [] (110bp,23bp) -- (110bp,65.99bp);
  \draw (121.5bp,45bp) node {$10$};
  \draw [] (68.555bp,144.18bp) -- (82.167bp,187.61bp);
  \draw (86.5bp,165bp) node {$47$};
\begin{scope}
  \definecolor{strokecol}{rgb}{0.0,0.0,0.0};
  \pgfsetstrokecolor{strokecol}
  \draw (20bp,23bp) -- (0bp,23bp) -- (0bp,2bp) -- (20bp,2bp) -- cycle;
  \draw (10bp,12bp) node {$R$};
\end{scope}
\begin{scope}
  \definecolor{strokecol}{rgb}{0.0,0.0,0.0};
  \pgfsetstrokecolor{strokecol}
  \draw (61bp,23bp) -- (39bp,23bp) -- (39bp,2bp) -- (61bp,2bp) -- cycle;
  \draw (50bp,12bp) node {$r_1$};
\end{scope}
\begin{scope}
  \definecolor{strokecol}{rgb}{0.0,0.0,0.0};
  \pgfsetstrokecolor{strokecol}
  \draw (91bp,23bp) -- (69bp,23bp) -- (69bp,2bp) -- (91bp,2bp) -- cycle;
  \draw (80bp,12bp) node {$r_2$};
\end{scope}
\begin{scope}
  \definecolor{strokecol}{rgb}{0.0,0.0,0.0};
  \pgfsetstrokecolor{strokecol}
  \draw (120bp,23bp) -- (100bp,23bp) -- (100bp,2bp) -- (120bp,2bp) -- cycle;
  \draw (110bp,12bp) node {$\theta_1$};
\end{scope}
\begin{scope}
  \definecolor{strokecol}{rgb}{0.0,0.0,0.0};
  \pgfsetstrokecolor{strokecol}
  \draw (155bp,23bp) -- (135bp,23bp) -- (135bp,2bp) -- (155bp,2bp) -- cycle;
  \draw (145bp,12bp) node {$\theta_2$};
\end{scope}
\begin{scope}
  \definecolor{strokecol}{rgb}{0.0,0.0,0.0};
  \pgfsetstrokecolor{strokecol}
  \draw (88bp,195bp) ellipse (9bp and 9bp);
  \draw (88bp,195bp) node {$$};
\end{scope}
\begin{scope}
  \definecolor{strokecol}{rgb}{0.0,0.0,0.0};
  \pgfsetstrokecolor{strokecol}
  \draw (68bp,135bp) ellipse (9bp and 9bp);
  \draw (68bp,135bp) node {$$};
\end{scope}
\begin{scope}
  \definecolor{strokecol}{rgb}{0.0,0.0,0.0};
  \pgfsetstrokecolor{strokecol}
  \draw (11bp,75bp) ellipse (9bp and 9bp);
  \draw (11bp,75bp) node {$$};
\end{scope}
\begin{scope}
  \definecolor{strokecol}{rgb}{0.0,0.0,0.0};
  \pgfsetstrokecolor{strokecol}
  \draw (68bp,75bp) ellipse (9bp and 9bp);
  \draw (68bp,75bp) node {$$};
\end{scope}
\begin{scope}
  \definecolor{strokecol}{rgb}{0.0,0.0,0.0};
  \pgfsetstrokecolor{strokecol}
  \draw (110bp,135bp) ellipse (9bp and 9bp);
  \draw (110bp,135bp) node {$$};
\end{scope}
\begin{scope}
  \definecolor{strokecol}{rgb}{0.0,0.0,0.0};
  \pgfsetstrokecolor{strokecol}
  \draw (110bp,75bp) ellipse (9bp and 9bp);
  \draw (110bp,75bp) node {$$};
\end{scope}
\begin{scope}
  \definecolor{strokecol}{rgb}{0.0,0.0,0.0};
  \pgfsetstrokecolor{strokecol}
  \draw (145bp,75bp) ellipse (9bp and 9bp);
  \draw (145bp,75bp) node {$$};
\end{scope}
\end{tikzpicture}
}
\subfigure{
\begin{tikzpicture}[>=latex,line join=bevel, scale=0.5, every node/.style={scale=0.96}]
  \pgfsetlinewidth{0.6bp}
\pgfsetcolor{black}

\draw(20bp,190bp) node {$1 \cdot 10^{-5}\,\text{eV}$};
\draw(20bp,165bp) node {12.6\%};

  \draw [] (16.46bp,82.427bp) -- (60.553bp,129.38bp);
  \definecolor{strokecol}{rgb}{0.0,0.0,0.0};
  \pgfsetstrokecolor{strokecol}
  \draw (55.5bp,105bp) node {$29$};
  \draw [] (145bp,23bp) -- (145bp,65.965bp);
  \draw (156.5bp,45bp) node {$10$};
  \draw [] (48.738bp,23bp) -- (61.177bp,68.646bp);
  \draw (61bp,45bp) node {$8$};
  \draw [] (10.17bp,23bp) -- (10.852bp,65.7bp);
  \draw (25bp,45bp) node {$192$};
  \draw [] (110bp,84.113bp) -- (110bp,125.8bp);
  \draw (121.5bp,105bp) node {$10$};
  \draw [] (144.06bp,84.218bp) -- (117.78bp,130.33bp);
  \draw (152.5bp,105bp) node {$10$};
  \draw [] (78.072bp,23bp) -- (69.728bp,65.925bp);
  \draw (84bp,45bp) node {$8$};
  \draw [] (106.85bp,143.6bp) -- (91.146bp,186.42bp);
  \draw (113.5bp,165bp) node {$89$};
  \draw [] (68bp,84.113bp) -- (68bp,125.8bp);
  \draw (79.5bp,105bp) node {$50$};
  \draw [] (110bp,23bp) -- (110bp,65.99bp);
  \draw (121.5bp,45bp) node {$10$};
  \draw [] (68.555bp,144.18bp) -- (82.167bp,187.61bp);
  \draw (86.5bp,165bp) node {$89$};
\begin{scope}
  \definecolor{strokecol}{rgb}{0.0,0.0,0.0};
  \pgfsetstrokecolor{strokecol}
  \draw (20bp,23bp) -- (0bp,23bp) -- (0bp,2bp) -- (20bp,2bp) -- cycle;
  \draw (10bp,12bp) node {$R$};
\end{scope}
\begin{scope}
  \definecolor{strokecol}{rgb}{0.0,0.0,0.0};
  \pgfsetstrokecolor{strokecol}
  \draw (61bp,23bp) -- (39bp,23bp) -- (39bp,2bp) -- (61bp,2bp) -- cycle;
  \draw (50bp,12bp) node {$r_1$};
\end{scope}
\begin{scope}
  \definecolor{strokecol}{rgb}{0.0,0.0,0.0};
  \pgfsetstrokecolor{strokecol}
  \draw (91bp,23bp) -- (69bp,23bp) -- (69bp,2bp) -- (91bp,2bp) -- cycle;
  \draw (80bp,12bp) node {$r_2$};
\end{scope}
\begin{scope}
  \definecolor{strokecol}{rgb}{0.0,0.0,0.0};
  \pgfsetstrokecolor{strokecol}
  \draw (120bp,23bp) -- (100bp,23bp) -- (100bp,2bp) -- (120bp,2bp) -- cycle;
  \draw (110bp,12bp) node {$\theta_1$};
\end{scope}
\begin{scope}
  \definecolor{strokecol}{rgb}{0.0,0.0,0.0};
  \pgfsetstrokecolor{strokecol}
  \draw (155bp,23bp) -- (135bp,23bp) -- (135bp,2bp) -- (155bp,2bp) -- cycle;
  \draw (145bp,12bp) node {$\theta_2$};
\end{scope}
\begin{scope}
  \definecolor{strokecol}{rgb}{0.0,0.0,0.0};
  \pgfsetstrokecolor{strokecol}
  \draw (88bp,195bp) ellipse (9bp and 9bp);
  \draw (88bp,195bp) node {$$};
\end{scope}
\begin{scope}
  \definecolor{strokecol}{rgb}{0.0,0.0,0.0};
  \pgfsetstrokecolor{strokecol}
  \draw (68bp,135bp) ellipse (9bp and 9bp);
  \draw (68bp,135bp) node {$$};
\end{scope}
\begin{scope}
  \definecolor{strokecol}{rgb}{0.0,0.0,0.0};
  \pgfsetstrokecolor{strokecol}
  \draw (11bp,75bp) ellipse (9bp and 9bp);
  \draw (11bp,75bp) node {$$};
\end{scope}
\begin{scope}
  \definecolor{strokecol}{rgb}{0.0,0.0,0.0};
  \pgfsetstrokecolor{strokecol}
  \draw (68bp,75bp) ellipse (9bp and 9bp);
  \draw (68bp,75bp) node {$$};
\end{scope}
\begin{scope}
  \definecolor{strokecol}{rgb}{0.0,0.0,0.0};
  \pgfsetstrokecolor{strokecol}
  \draw (110bp,135bp) ellipse (9bp and 9bp);
  \draw (110bp,135bp) node {$$};
\end{scope}
\begin{scope}
  \definecolor{strokecol}{rgb}{0.0,0.0,0.0};
  \pgfsetstrokecolor{strokecol}
  \draw (110bp,75bp) ellipse (9bp and 9bp);
  \draw (110bp,75bp) node {$$};
\end{scope}
\begin{scope}
  \definecolor{strokecol}{rgb}{0.0,0.0,0.0};
  \pgfsetstrokecolor{strokecol}
  \draw (145bp,75bp) ellipse (9bp and 9bp);
  \draw (145bp,75bp) node {$$};
\end{scope}
\end{tikzpicture}
}
\caption{MLPF trees with truncation ranks (upper two layers) and primitive
basis sizes (bottom layer) for the zeroth Fourier component of the
BMKP potential, for different values of the target $\Delta_\text{rms}$.  The percentages
indicate the compression ratio (see text).}
\label{fig:mlpf-tree}
\end{figure}

The aim of this example study was to investigate how the accuracy of
this truncation affects the computational effort and the results. From \refsec{sec:numeff}
it is expected that the computational advantage of MLPF over Potfit will be
larger for higher accuracies. To verify this expectation, both Potfit and MLPF
representations of the Fourier components $V_0$, \ldots, $V_3$ were created for
target values of $\Delta_\text{rms}$ ranging from $2\cdot 10^{-3}$ to $1\cdot 10^{-5}\,\text{eV}$.
For Potfit, the mode $(r_1,r_2)$ was contracted, and the numbers of natural potentials for the
other modes were manually adjusted to reach the desired accuracy. For MLPF,
the truncation ranks were obtained through the error-controlled truncation
algorithm described in  \refapp{app:errctrl}.
In all cases, the upper bound for the actual $\Delta_\text{rms}$, as returned
by the algorithm, was quite close to the supplied target value. 
The resulting MLPF trees for $V_0$ are shown in \reffig{fig:mlpf-tree}, where also the
truncation ranks required to achieve the wanted accuracy as well as the ``compression ratio''
are indicated.  The latter measures how much data the MLPF representation needs
compared to the original data given on the primitive product grid.
(Other Fourier components show similar behaviour.) 

\begin{figure}
\includegraphics[scale=1]{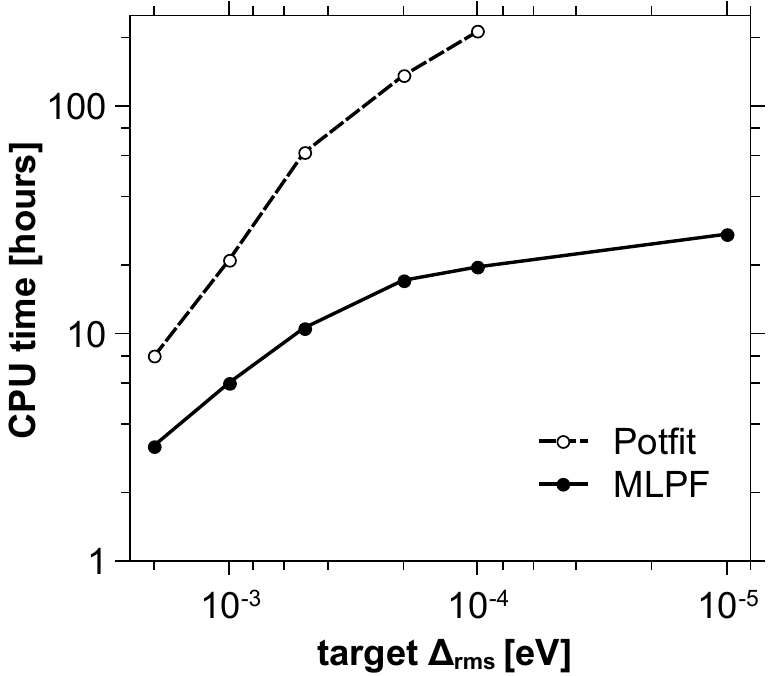}
\caption{CPU time needed for an ML-MCTDH propagation versus
the accuracy of the potential representation, for a system with 4 primitive modes.
(See text for details of the system.)  The open and filled circles show the timings for
the potential in Potfit and MLPF representation, respectively. Data was obtained
using one core of an Intel$^\text{\textregistered}$ i5-3470 CPU.}
\label{fig:timing-bmkp}
\end{figure}

\begin{table}
\begin{tabular}{rcrccrcrc}
\hline\hline
\multirow{3}{*}{\raisebox{1.3em}{\parbox[b]{\widthof{$\Delta_\text{rms}$ [eV]}}{target\\$\Delta_\text{rms}$ [eV]}}} &
\phantom{abc} &
\multirow{3}{*}{\raisebox{1.3em}{\parbox[b]{\widthof{term count}}{Potfit\\term count}}} &
\phantom{abc} &
\multicolumn{5}{c}{memory usage [MiB]}
\\
\cline{5-9}
&&&&
\phantom{ab} &
Potfit &
\phantom{ab} &
MLPF &
\phantom{ab}
\\
\hline
$2 \cdot 10^{-3}$ &&   300 &&&   73 &&   55 \\
$1 \cdot 10^{-3}$ &&   800 &&& 121 &&   63 \\
$5 \cdot 10^{-4}$ && 2299 &&& 263 &&   74 \\
$2 \cdot 10^{-4}$ && 5144 &&& 533 &&   90 \\
$1 \cdot 10^{-4}$ && 8044 &&& 807 && 105 \\
$1 \cdot 10^{-5}$ &&           &&&         && 135 \\
\hline\hline
\end{tabular} 
\caption{Memory consumption of an ML-MCTDH propagation 
for a system with 4 primitive modes (see text for details),
using a potential either in Potfit-format or in MLPF-format.
For Potfit, the memory usage is close to proportional to the
number of terms in the sum-of-products expansion.}
\label{tab:memusage}
\end{table}

With the potential in either Potfit- or MLPF-format, an ML-MCTDH propagation
was carried out using an extended version of the Heidelberg MCTDH package%
\cite{mctdh85} which can evaluate the ML-MCTDH EOMs for multi-layer operators.
The initial state is described by the two molecules in their vibrational and rotational
ground state, with the relative motion described by a Gaussian wavepacket centered
at $R=8.0\,a_0$ with a width of $0.3\,a_0$ and a momentum of $8.0\,\hbar a_0^{-1}$,
at zero total angular momentum. After propagating for $250\,\text{fs}$, the method
of Tannor and Weeks\cite{tan93:3884} was used to extract
energy-resolved transition probabilities $P_{i \to f}(E)$ for
rotational excitations of the molecules from the time-dependent wavefunction
(see \refcite{ott08:064305} for details of this analysis procedure). Here, only the
four strongest rotational transitions from $i=(j_1,j_2)=(0,0)$ to $f=(j'_1,j'_2)=(2,0), (2,2), (4,0), (4,2)$
were considered (the $j_i$ are the rotational quantum numbers of the molecules;
changes in the vibrational states were not considered).

\reffig{fig:timing-bmkp} plots the CPU time needed for one such ML-MCTDH propagation
against the target accuracy of the potential representation. (For the present system,
the CPU time for producing the Potfit- and MLPF-format potentials are negligible.)
As predicted, the MLPF representation shows a significantly weaker increase of the numerical effort with
increasing accuracy than the Potfit representation. The computational savings factor
ranges from $2.5$ to $10.8$, which roughly confirms the estimate given in \refsec{sec:numeff}.
Likewise, Table \ref{tab:memusage} summarizes the peak memory consumption for the
ML-MCTDH propagations. Again, the resource usage is strongly reduced when using the
potential in MLPF-format.  For the target $\Delta_\text{rms} = 10^{-5}\,\text{eV}$, a propagation using
Potfit-format would have needed a prohibitive amount of time for the present study, and only
the propagation using MLPF-format was carried out; its results were used as a reference to judge
the actual accuracy of the other propagations.

\begin{figure}
\includegraphics[scale=1]{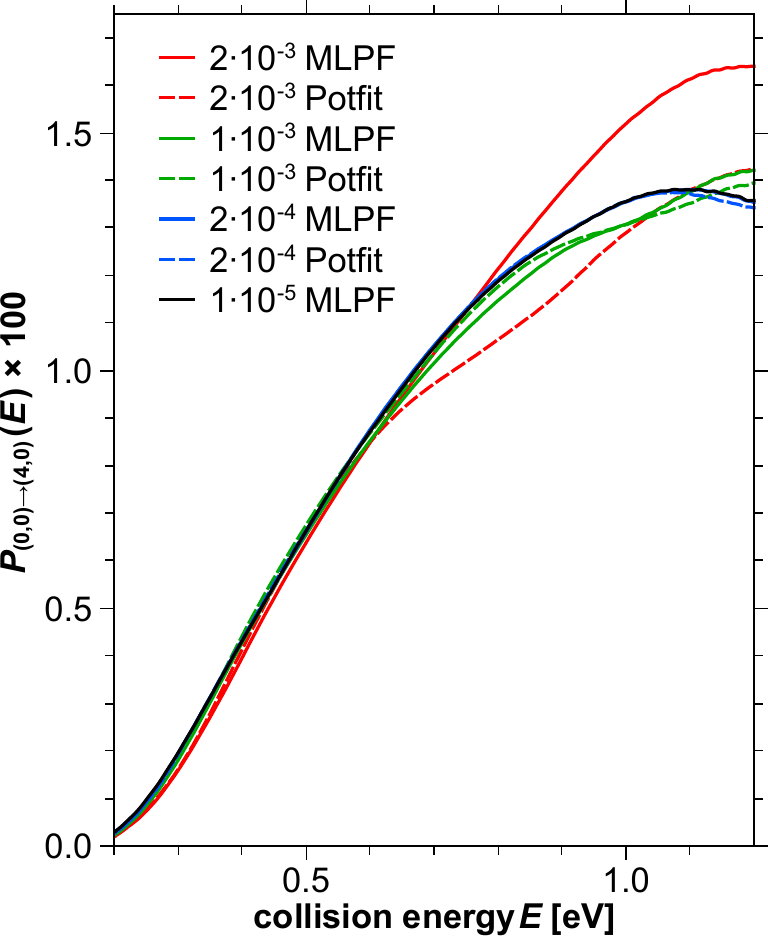}
\caption{(color online) Convergence of the transition probability
$P_{(0,0)\to(4,0)}$ with increasing accuracy of the potential representation.
The legend lists the target accuracy $\Delta_\text{rms}$ (in eV) and the
representation method.  This transition was selected because it exhibits
the largest deviations. Compare also with Table \ref{tab:accuracy}.}
\label{fig:tprob}
\end{figure}

While Potfit and MLPF yield approximations to the potential with
a controllable RMS error, the question is how the physically relevant
observables are affected by these approximations. To verify that
more accurate potentials indeed lead to more accurate observables,
the computed transition probabilities $P_{i \to f}$ were compared to the results
of the reference propagation (MLPF with $\Delta_\text{rms} = 10^{-5}\,\text{eV}$).
Among the rotational transitions considered, the one with the final state
$f=(4,0)$ exhibited the most prominent differences. The convergence of
the corresponding transition probabilities is displayed in \reffig{fig:tprob}.
One can see that both Potfit and MLPF exhibit similar convergence behaviour.
To make a more quantifiable statement, the following error measure can be used:
\begin{align}
\text{$P$-RMSE} = \sqrt{
\frac{  \sum\limits_{i \to f} \, \int_{E_0}^{E_1} \! \text{d}E
\left( P_{i \to f}(E) - P^\text{ref}_{i \to f}(E) \right)^2 }%
{\sum\limits_{i \to f}^{\:} (E_1-E_0)}}
\end{align}
where the integration ranges over the collision energy from
$E_0=44\,\text{meV}$ (the energy threshold of the lowest
rotational transition) to $E_1=1.2\,\text{eV}$,
and the sum includes the four strongest rotational transitions. Table \ref{tab:accuracy}
shows that this error measure indeed roughly follows the accuracy of the potential
representation, and that Potfit and MLPF lead to similar errors in the observables.
This indicates that Potfit and MLPF lead to the same results, within  the
specified margin of accuracy.

\begin{table}
\begin{tabular}{rccrcrc}
\hline\hline
\multirow{3}{*}{\raisebox{1.3em}{\parbox[b]{\widthof{$\Delta_\text{rms}$ [eV]}}{target\\$\Delta_\text{rms}$ [eV]}}} &
\phantom{abc} &
\multicolumn{5}{c}{$P$-RMSE [$10^{-3}$]}
\\
\cline{3-7}
&&
\phantom{ab} &
Potfit &
\phantom{ab} &
MLPF &
\phantom{ab}
\\
\hline
$2 \cdot 10^{-3}$ &&& 3.27 && 3.33 &\\
$1 \cdot 10^{-3}$ &&& 1.39 && 1.98 &\\
$5 \cdot 10^{-4}$ &&& 1.21 && 0.51 &\\
$2 \cdot 10^{-4}$ &&& 0.26 && 0.24 &\\
$1 \cdot 10^{-4}$ &&& 0.23 && 0.18 &\\
\hline\hline
\end{tabular}
\caption{Error measure (see text) of the transition probabilities $P_{i \to f}$,
summed over the four strongest rotational transitions, using a potential
either in Potfit- or MLPF-format.  For comparison, the transition probabilities here
range between $0.0$ and $0.28$, with an average value around $0.09$.}
\label{tab:accuracy}
\end{table}

\section{Summary and Outlook}
\label{sec:summary}

While multi-layer MCTDH (ML-MCTDH) has already been successfully used
for several high-dimensional systems%
\cite{wan03:1289,cra07:144503,wan10:78,wes11:184102,cra11:064504,ven11:044135,men12:134302,men13:014313},
it has so far been difficult to use it
for systems that are described by a general potential energy surface (PES).
This is due to the difficulties encountered when evaluating matrix elements
of the potential energy operator. The CDVR approach%
\cite{man96:6989,man09:054109} has been successful
in overcoming these difficulties%
\cite{ham11:224305,ham12:054105,wes11:184102,wod12:11249,wes13:014309,wel12:244106,wel13:164118},
though its use requires an enormous amount
of PES evaluations, which may require a large amount of computational
resources.  The alternative to CDVR is to represent (or fit) the PES in a form
that is compatible with ML-MCTDH, in the sense that it allows efficient evaluation
of the matrix elements. Traditionally, the sum-of-products form is used, and
the fitting of the PES into this form is done by the Potfit algorithm\cite{jae96:7974}. However,
for large systems the Potfit representation requires a huge amount of terms
so that ML-MCTDH computations may become impractical.

This article has introduced the multi-layer Potfit (MLPF) method, which
yields a PES representation that is as accurate as Potfit, that is
as computationally affordable as Potfit, and that can reduce the computational
cost for ML-MCTDH by orders of magnitude. Technically, this is achieved
by employing the hierarchical singular value decomposition\cite{gra10:2029}
to represent the PES in the hierarchical tensor format\cite{hac09:706} (Section \ref{sec:mlpf}).
This procedure not only yields a near-optimal fit, its accuracy can also
easily be estimated on the fly.  Using these features, the present work
shows that this algorithm can be used in a black-box-like manner, in the
sense that all expansion orders could be automatically selected based on
the desired root-mean-square error of the representation.

The MLPF representation of the PES allows for an efficient recursive
scheme for evaluating all the ML-MCTDH matrix elements of the potential
energy operator (Section \ref{sec:mlop+mlmctdh}). The computational cost
for this scheme has been estimated and compared to the cost induced by the
Potfit representation (Section \ref{sec:numeff}). This estimate shows that
MLPF can already be beneficial for systems with just four primitive
modes, where computational savings of about one order of magnitude
can be expected. In general, the savings are higher for more accurate
representations, so that MLPF can actually yield more accurate results
with less computational effort than Potfit. For a larger number of
primitive modes, by using MLPF instead of Potfit, the computational
effort for ML-MCTDH is expected to decrease by orders of magnitude.
This means that computations once considered infeasible may now be
within reach.

To test the methods discussed in this paper, an implementation of the
hierarchical SVD algorithm has been written in Fortran 2003, and the
Heidelberg MCTDH package\cite{mctdh85} has been extended so that
it can use the required multi-layer operator structure and
treat the modified equations of motion. The code was used
to study a diatom-diatom scattering system with seven degrees of freedom
that are organized into four primitive modes (Section \ref{sec:compex}).
Representations of the PES in Potfit- and in MLPF-format
were created for different target accuracies, and an ML-MCTDH propagation
with each of these PES representations was carried out. It was confirmed
that Potfit and MLPF lead to the same results within the given margin
of accuracy. The computational effort using the MLPF-form is indeed 
much lower than for the Potfit-form, especially for high-accuracy representations,
where the speedup was more than a factor of ten. These results fully confirm
the theoretical estimates.

The biggest limitation for MLPF is that it currently, like Potfit, needs
to initially evaluate the PES on the full product grid. This limits its applicability
to systems with 6--8 degrees of freedom (4--5 primitive modes). The recently
introduced multigrid Potfit (MGPF) method\cite{pel13:014108} can relax this
limitation, and is easy to integrate with MLPF. This should allow treating
systems with up to 12 degrees of freedom, and an example study (to be
published separately) of a 9D system has so far yielded encouraging results.
For larger numbers of dimensions, one must likely develop new methods
for finding a good MLPF representation of the PES. As a starting point, it is
worth noticing that MLPF is quite agnostic regarding the grid structure within the
primitive modes. Currently, ML-MCTDH (and MLPF) can employ multi-dimensional
primitive modes but requires them to use a direct product grid, so that
already 3D modes become very large. Instead, one might use optimized
multi-dimensional non-product grids, which will bring the modes down
to a more manageable size, thus enabling the treatment of systems with
a few additional degrees of freedom. Additionally, the properties of the
hierarchical SVD, like giving a strict upper bound for the error, are almost
\emph{too} good -- more approximate methods that can still work for
larger systems will probably be acceptable in practice.

\begin{acknowledgments}

The author wishes to thank Hans-Dieter Meyer and Daniel Peláez
for carefully reading and commenting on the manuscript. Also, the
author is grateful to them as well as to Oriol Vendrell
for encouraging discussions.
Financial support by the Deutsche Forschungsgemeinschaft
(DFG) is gratefully acknowledged.

\end{acknowledgments}

\appendix

\section{Technical details of multi-layer Potfit}
\label{app:mlpftech}

\subsection{The MLPF algorithm}
\label{app:mlpf}

Procedures for fitting full tensors into the hierarchical tensor format
have been described by Grasedyck in \refcite{gra10:2029} and termed
\emph{hierarchical singular value decomposition}. The computationally
most affordable of these algorithms is the one using \emph{leaves-to-root truncation}
(ibid., Algorithm 2). However, \refcite{gra10:2029} (as well as other mathematical
literature) only considers binary trees, whereas the trees used with ML-MCTDH
may have nodes with more than two children. Hence, a minor extension of the
algorithm is needed.

In the case that all leaf nodes of the ML-tree are on the same layer
(i.e. they have they same distance from the top node),
the MLPF algorithm proceeds as follows:
\begin{enumerate}
\item
Obtain the potential tensor $V_{\alpha_1 \cdots \alpha_d}$.
\item
Use the Potfit/HOSVD algorithm\cite{jae96:7974,lat00:1253} on $V$: First,
for each leaf node $z=\prim{f}$ \ldots
\begin{enumerate}
\item
rearrange $V$ into a matrix, with $\alpha_f$ as the row index
and all other indices combined into the column index
\item
perform a SVD of this matrix $\Rightarrow$ singular values $\sigma^\prim{f}_i$
and left singular vectors $v^\prim{f}_i$
\item
keep only the $m_z$ dominant singular vectors
\end{enumerate}
Then compute the core tensor $C$ by \refeq{eq:pfcore}.
\item
Move up one layer in the tree.
If the top node is reached, the algorithm is finished.
Otherwise, let there be $d'$ nodes in this layer.
For each node $z$ in this layer, combine the indices $i_{z,1},\ldots,i_{z,p_z}$
of its children into a multi-index $J^z$, so that the core tensor
$C$ can be reshaped into an order-$d'$ tensor.
\item
Perform HOSVD on the reshaped $C$. This yields, for each $z$ on the current
layer, singular values $\sigma^{(z)}_i$ (and the associated natural
weights, $\lambda^{(z)}_i := [\sigma^{(z)}_i]^2$) and the $m_z$ dominant singular vectors
$v^{(z)}_i$. Update the core tensor by projecting the current core tensor onto
outer products of the singular vectors, analogous to \refeq{eq:pfcore}.
\item
Go to step 3.
\end{enumerate}
The more general case, where leaf nodes may be on different layers,
can be dealt with by treating the corresponding indices of the core tensor
as ``by-standers'', i.e. no SVD and no projection is done along those
indices, until the computation has progressed so far that the the node
in question is on the current layer. 

\subsection{Error-controlled truncation}
\label{app:errctrl}

The error caused by MLPF can be estimated via \refeq{eq:hiersvd-error}.
Because the natural weights are available while the algorithm progresses,
one can devise a scheme which automatically chooses the truncation
ranks $m_z$ such that the MLPF fit will have a specified accuracy.
In the present work, the following scheme is used:
\begin{enumerate}
\item
Select the desired root-mean-square (RMS) error $\Delta_\text{rms}$ of the
MLPF representation. Set $\Delta^2 = N_\text{grid} \Delta_\text{rms}^2$,
where $N_\text{grid}$ is the number of points in the full product grid. Set
$K_\text{todo} = K-1$. Set the current layer to the lowest layer.
\item
Set $\epsilon^2 = \Delta^2 / K_\text{todo}$. Perform the HOSVD of the current
layer, choosing truncation ranks $m_z$ such that $\sum_{i>m_z} \lambda^{(z)}_i \leq \epsilon^2$
for all $z$ on the current layer.
\item
Update $\Delta^2$ by subtracting from it the sum of all neglected natural weights.
Update $K_\text{todo}$ by subtracting from it the number of nodes on the current layer.
\item
Go up one layer. If the top node is reached, exit. If there are exactly two nodes on the
current layer, reduce $K_\text{todo}$ from 2 to 1 (because the next HOSVD will be
a regular matrix SVD). Go to step 2.
\end{enumerate}
Choosing the truncation ranks in this way ensures that the resulting potential
approximation will have an RMS error that is not higher than $\Delta_\text{rms}$.

The above scheme tries to distribute the error evenly across all nodes, but if
nodes on lower layers don't use up the errors that are allocated to them, nodes
on higher layers are allowed to use larger errors. This may be beneficial because
it is expected that nodes closer to the top need larger truncation ranks, so this
scheme helps to keep the truncation ranks more uniform.
Admittedly, the given scheme is quite ad-hoc; other schemes are conceivable,
e.g. one may try to distribute the error evenly across layers. 

\section{Analysis of computational cost for ML-MCTDH}
\label{app:numeff}

\subsection{ML-MCTDH with sum-of-products operators}
\label{app:numeff-pf}

Let us estimate the numerical effort that is required to evaluate the
ML-MCTDH EOMs with operators in sum-of-products format
(see \refsec{sec:mlmctdh-sop}). The strategy is to first compute all
$\mathfrak{h}$-terms by \refeq{eq:hterms}, then all $\mathfrak{H}$-terms
by \refeq{eq:mfterms}, and finally the RHS of Eqs.\,(\ref{eq:eompf-top}--\ref{eq:eompf-leaf}).
To illustrate how to estimate the numerical effort of these computations,
take as an example the $\mathfrak{h}$-terms for a non-leaf
node $z$ with $p_z=3$, $n_z=\bar{n}$, and $n_{z,\kappa}=n$, so that
\refeq{eq:hterms} reads
\begin{align}
\mathfrak{h}^{(z)}_{r,jk} &=
\sum_{i_1=1}^n \sum_{i_2=1}^n \sum_{i_3=1}^n 
\sum_{l_1=1}^n \sum_{l_2=1}^n \sum_{l_3=1}^n 
A^{(z)*}_{j;i_1 i_2 i_3} A^{(z)}_{k;l_1 l_2 l_3}
\nonumber\\&\quad\times\:
\mathfrak{h}^{(z,1)}_{r,i_1 l_1} \mathfrak{h}^{(z,2)}_{r,i_2 l_2} \mathfrak{h}^{(z,3)}_{r,i_3 l_3}
\quad( j,k = 1 \ldots \bar{n} )
\:.
\end{align}
In mathematical terms, $A^{(z)}_{k;l_1 l_2 l_3}$ is a tensor of order 4, and
(considering only a single $r$) the $\mathfrak{h}^{(z,\kappa)}$ are matrices,
and the summations over the $l_\kappa$ constitute matrix-tensor
multiplications which can be carried out successively. The first such
operation over, say, $l_1$ produces from $A^{(z)}$ and $\mathfrak{h}^{(z,1)}$ an intermediate order-4 tensor
$X'_{k;i_1 l_2 l_3}$. This tensor has $\bar{n}n^3$ elements, and each element
takes $n$ FLOPs (combined floating-point multiplications and additions) to
compute, leading to an overall cost of $\bar{n}n^4$.
\footnote{While the scaling of this cost might theoretically be reduced by using
matrix-multiplication algorithms like Strassen's, the values of $n,\bar{n}$ are
in practice much too small for this to be worthwhile.}
In the same manner, the summation over $l_2$ produces from $X'$ and $\mathfrak{h}^{(z,2)}$
another intermediate tensor $X''_{k;i_1 i_2 l_3}$, which costs another
$\bar{n}n^4$ FLOPs. Likewise for the summation over $l_3$, which produces
a tensor $X'''_{k;i_1 i_2 i_3}$.  Finally, the order-4 tensors $A^{(z)*}$
and $X'''$ are contracted over the indices $i_1,i_2,i_3$ to produce the final
matrix $\mathfrak{h}^{(z)}$. This matrix has $\bar{n}^2$ elements, and
each contraction costs $n^3$ FLOPs, resulting in an overall cost of $\bar{n}^2n^3$
for this last step. Hence the total cost for computing $\mathfrak{h}^{(z)}_{r,jk}$
for one $r$ and all $j,k$ is, in this specific case, $\bar{n}n^3(\bar{n}+3n)$, or more
generally for $p_z=p$, $\bar{n}n^p(\bar{n}+pn)$. 
Finally, computing these terms for all $r$, i.e. for all terms in the sum-of-products
expansion (\ref{eq:sumprod}), increases the overall cost by a factor of $s$, i.e.
the number of terms in this expansion.

In a similar manner, one may estimate the computational effort for
the $\mathfrak{H}$-terms (\ref{eq:mfterms}) and for the RHS of the
EOMs (\ref{eq:eompf-top}--\ref{eq:eompf-leaf}). In case of the
latter, the effort for multiplying with the inverse density matrix and for
applying the projector can usually be neglected, as these need to be done only
once so that their cost doesn't depend on $s$. In summary,
for a non-leaf node $z$ with $p_z=p$, $n_z=\bar{n}$, and $n_{z,\kappa}=n$,
one obtains:
\begin{align}
\text{cost}(\mathfrak{h}^{(z)}) &= s\bar{n}n^p(\bar{n}+pn)
\label{cost:hterms-pf}
\\
\text{cost}(\mathfrak{H}^{(z,1)}) + \cdots + \text{cost}(\mathfrak{H}^{(z,p)}) &= s\bar{n}n^p(\bar{n}+p^2n)
\label{cost:mfterms-pf}
\\
\text{cost}(\text{RHS}^{(z)}) &\approx s\bar{n}n^p(\bar{n}+pn)
\label{cost:rhs-pf}
\end{align}
For leaf nodes, the computational effort depends on the sparsity of the matrix
representations of the primitive mode operators,
$\langle \chi^\prim{f}_\alpha | \hat{h}^\prim{f}_r | \chi^\prim{f}_\beta \rangle$.
If this matrix is diagonal, the effort will be proportional to $N_f$; for a
full matrix, it is $\propto N_f^2$; if the primitive mode employs
an FFT representation, the effort is $\propto N_f \log N_f$.
But as was shown in section \ref{sec:potfit}, if a general PES in
Potfit form is used, most of the terms $\hat{h}^\prim{f}_r$ have a diagonal
matrix representation, and to simplify the discussion, let us assume that
this is true for \emph{all} terms. Then the computational effort for 
a leaf node $z=\prim{f}$ with $n_z=\bar{n}$ and $N_f=N$ becomes
\begin{align}
\text{cost}(\mathfrak{h}^{(z)}) &= s\bar{n}N(\bar{n}+1)
\label{cost:hterms-pf-prim}
\\
\text{cost}(\text{RHS}^{(z)}) &\approx s\bar{n}N(\bar{n}+1)
\label{cost:rhs-pf-prim}
\end{align}
Note that there is no equivalent of \refeq{cost:mfterms-pf} for leaf
nodes because they don't have children.

To illustrate the total numerical effort for the \emph{whole} ML-tree,
a homogeneous tree structure like in \refsec{sec:numeff} is assumed.
But here the restriction to binary trees is lifted, and more general
trees with mode combinations of order $p$ are considered,
so that there are $d=p^L$ primitive modes in the tree. As in \refsec{sec:numeff},
the number of SPFs for each primitive mode is assumed to be $n$,
and increases by a  factor of $a$ when going one layer up.
Then, summing all relevant costs from Eqs.\,(\ref{cost:hterms-pf}--\ref{cost:rhs-pf-prim})
over all nodes, and keeping only the most expensive terms, results in:
\begin{align}
\text{wf-size} &\propto n^{p+1} d^{(p+1) \log_p a}
\quad(\text{for } a>p^{1/(p+1)})
\\
\text{total-cost} &\propto s\,n^{p+2} d^{(p+2) \log_p a}
\quad(\text{for } a>p^{1/(p+2)})
\label{eq:totcost-sop}
\end{align}
For $p=2$, this reproduces Manthe's result\cite{man08:164116} for the size of the ML-MCTDH wavefunction,
but the numerical effort shows a slightly worse scaling behaviour, namely $s\,n^4 d^{4\,log_2 a}$.
By practical experience (mostly with primitive mode combination in standard MCTDH)
it was found that $a=p$ is a reasonable value, which leads  to a polynomial dependence
of the numerical effort on the number of degrees of freedom ($d^4$ for $p=2$.). However, this
result is negated if $s$ has a stronger dependence on $d$, which happens to be the case if
Potfit is used to represent the potential.

\subsection{ML-MCTDH with multi-layer operators}
\label{app:numeff-mlpf}

Let us now estimate the numerical effort for the ML-MCTDH equations of motion in case
a multi-layer operator is used. Again, the overall strategy
is to first evaluate the $\mathfrak{U}$-terms (\ref{eq:uterms-leaf},\ref{eq:uterms}),
then the $\mathfrak{W}$-terms (\ref{eq:wterms}), and finally the RHS contributions
(\ref{eq:eom-mlpf-top}--\ref{eq:eom-mlpf-leaf}). Naturally, the analysis of the computational costs
for these expressions is now more involved.

For example, computing the $\mathfrak{U}$-terms for an internal node $z$
with $p_z=p$ involves an expression (cf. \refeq{eq:uterms}) with 3 tensors of order $p+1$
($V^{(z)}$, $A^{(z)*}$, $A^{(z)}$) and $p$ tensors of order 3 ($\mathfrak{U}^{(z,1)}, \ldots, \mathfrak{U}^{(z,p)}$)
which need to be contracted over $3p$ indices. The computational cost for this evaluation
depends on the order in which the contractions are carried out, and one may attempt to
find an order which minimizes this cost. This constitutes a generalization of the well-known
matrix-chain multiplication problem; although the generalization to arbitrary tensors has been
shown to be NP-complete\cite{lam97:157}, in the present case the number of tensors and
contractions is small enough to make an exhaustive search feasible. Using e.g. the
``single-term optimization'' algorithm from \refcite{har05:techrep}, this search can be
performed, and the results can guide the implementation.
Not surprisingly, the optimal order of contractions depends on the extents of the indices involved.
Moreover, the solution which minimizes the number of arithmetic operations may not
be optimal in practice, as it might require an excessive amount of temporary storage
(especially for larger $p_z$), and reducing memory consumption may anyway be beneficial
on contemporary computer architectures which employ a hierarchical memory model.
By considering ranges of values for $p_z$, $n_z$, $m_z$ which should be relevant in
practice, an approach was chosen that can be implemented rather generally, that
is not too far from optimal concerning computational cost, and that has moderate
requirements for temporary storage.

To illustrate this approach, let us consider the $\mathfrak{U}$-terms
for an internal node $z$ with $p_z=2$, $n_z=\bar{n}$, $m_z=\bar{m}$, $n_{z,\kappa}=n$, and
$m_{z,\kappa}=m$. These terms read (cf. \refeq{eq:uterms})
\begin{align}
\mathfrak{U}^{(z)}_{c,jk} &= \sum_{b_1=1}^{m} \sum_{b_2=1}^{m}
    \sum_{i_1=1}^{n} \sum_{i_2=1}^n \sum_{l_1=1}^n \sum_{l_2=1}^n
    V^{(z)}_{c;b_1 b_2}
\nonumber\\&\qquad\times\:
    A^{(z)*}_{j;i_1 i_2} A^{(z)}_{k;l_1 l_2}
    \mathfrak{U}^{(z,1)}_{b_1,i_1 l_1} \mathfrak{U}^{(z,2)}_{b_2,i_2 l_2}
\\
&\quad(c = 1 \ldots \bar{m}
\:;\: j,k = 1 \ldots \bar{n})
\:.
\nonumber
\end{align}
First, $A^{(z)}$ is contracted with $\mathfrak{U}^{(z,1)}$ over $i_1$
as well as $A^{(z)*}$ with $\mathfrak{U}^{(z,2)}$ over $l_2$ (cost:
$2 m \bar{n} n^3$). The two resulting intermediate tensors are then
contracted over $i_2,l_1$ (cost: $m^2 \bar{n}^2 n^2$). The resulting
tensor is finally contracted with $V^{(z)}$ over $b_1,b_2$ (cost:
$\bar{m} m^2 \bar{n}^2$). The evaluation of the $\mathfrak{W}$-terms
and the RHS contributions proceeds in a similar fashion, and the
required steps have the same computational costs as those encountered
for the $\mathfrak{U}$-terms.

To arrive at a cost estimate for the whole tree, consider the same
homogeneous tree structure as before, but for brevity now restricted to the case
$p=2$, i.e.  let all $d=2^L$ leaf nodes employ
$n$ SPFs and $m$ SPOs, and when going one layer up, let the number of SPFs
and SPOs increase by a factor of $a$ and $b$, respectively (except for
$n_\topn = m_\topn = 1$ at the top). In general, the cost for the bottom layer
can be neglected. Keeping only the highest-order terms in $n$ and $m$, one
obtains (using $x^L = d^{\log_2 x}$):
\begin{align}
&\text{node-cost}(\bar{n},n,\bar{m},m) \approx
c_1 m^2 \bar{n}^2 n^2 + c_2 \bar{m} m^2 \bar{n}^2
\\
&\phantom{\text{node-cost}(\bar{n},n,\bar{m},m) \approx}
\text{(with $c_1$, $c_2$ constant)}
\nonumber\\
&\text{total-cost} \approx
    \text{node-cost}(1, a^{L-1} n, 1, b^{L-1} m)
\nonumber\\
&\phantom{\text{total-cost}} +\:
    \sum_{l=1}^{L-1} 2^{L-l}\,\text{node-cost}(a^l n, a^{l-1} n, b^l m, b^{l-1} m)
\nonumber\\
&\quad\to c_1' m^2 n^4 (a^4 b^2)^L \,+\, c_2' m^3 n^2 (a^2 b^3)^L
\nonumber\\
&\mkern44mu
(\text{for } L\to\infty \text{ and if } a^4b^2,a^2b^3>2)
\nonumber\\
&\quad\to \begin{cases}
c_1' m^2 n^4 d^{4 \log_2 a + 2 \log_2 b} & \text{if } a^2 > b
\\
c_2' m^3 n^2 d^{2 \log_2 a + 3 \log_2 b} & \text{if } a^2 < b
\end{cases}
\label{eq:totcost-mlpf}
\end{align}
Assuming that $a$ and $b$ have similar values, the case
$a^2 > b$ is likely to be more relevant in practice, therefore
the discussion \refsec{sec:numeff} is restricted to that case.

\section*{References}

\end{document}